\documentclass{pasj01}

\begin{document}

\author{Mariko \textsc{Kimura}\altaffilmark{1,2,*},
        Yoji \textsc{Osaki}\altaffilmark{3},
        and Taichi \textsc{Kato}\altaffilmark{1}
        }
\email{mariko.kimura@riken.jp}

\altaffiltext{1}{Department of Astronomy, Graduate School of Science, Kyoto University, Oiwakecho, Kitashirakawa, Sakyo-ku, Kyoto 606-8502}
\altaffiltext{2}{Extreme Natural Phenomena RIKEN Hakubi Research Team, Cluster for Pioneering Research, RIKEN, 2-1 Hirosawa, Wako, Saitama 351-0198}
\altaffiltext{3}{Department of Astronomy, School of Science, University of Tokyo, Hongo, Tokyo 113-0033}

\title{
KIC 9406652: A laboratory of the tilted disk in cataclysmic variable stars
}

\Received{} \Accepted{}


\SetRunningHead{Kimura et al.}{}

\maketitle

\begin{abstract}

KIC 9406652 is a cataclysmic variable, sub-classified 
as `IW And-type star', showing a repetition of standstills 
with oscillatory variations terminated by brightening.  
\textcolor{black}{This system showed negative superhumps, 
semi-periodic variations having periods slightly shorter than 
the $\sim$6-hrs orbital period, and super-orbital signals 
having $\sim$4-d periods, both of which are believed to 
originate from a precessing, tilted accretion disk.}  
We have re-examined its {\it Kepler} light curve extending 
over 1500 d.  
In accordance with a cycle of the IW And-type light variation, 
the frequency of negative superhumps showed 
a reproducible variation: a rapid drop during the brightening 
and a gradual increase during the standstill.  
They are interpreted as the drastic change in the radial mass 
distribution and the expansion of the tilted disk, 
\textcolor{black}{which is not expected from the existing models 
for IW And stars}.
\textcolor{black}{The constancy in flux amplitudes} of 
negative superhumps confirms that their light source is 
the bright spot sweeping across the surface of the tilted disk.  
The frequencies of negative superhumps and super-orbital signals 
varied in unison on long timescales, suggesting their common origin: 
the tilted disk.  
\textcolor{black}{Orbital signals} at the brightening 
\textcolor{black}{were} dominated by the irradiation of 
the secondary star and varied with the orientation of 
the tilted disk; the amplitude was maximized 
at the minimum of super-orbital signals and the light 
maximum shifted to early orbital phases as 
the super-orbital phase advances.  
This is the first direct evidence that the disk was tilted 
out of the binary orbital plane and retrogradely precessing.
The tilt angle of the disk inferred from semi-amplitudes of 
super-orbital signals was lower than 3 degrees.  
The diversity in \textcolor{black}{light curves} of 
negative superhumps supports this and suggests that 
a part of the gas stream overflows the disk edge.  
This study thus offers rich information about the tilted disk 
in cataclysmic variables.

\end{abstract}

\section{Introduction}

Cataclysmic variables (CVs), semi-detached close binary 
systems consisting of a white dwarf (WD) (the primary star) 
and a low-mass cool star (the secondary star), are a group of 
eruptive variable stars to which novae and dwarf novae belong.  
An accretion disk is formed around the primary WD 
by the Roche-lobe overflow.  
The orbital period typically ranges from a few to several 
hours (\cite{war95book} for a general review).  
Dwarf novae (DNe) show intermittent outbursts with 
a typical amplitude between 2--5 mag in the optical band 
with an interval of a few weeks to hundred days and their 
outbursts are now believed to be caused by the thermal 
limit-cycle instability in the accretion disk 
(\cite{osa96review} for a review on the disk instability).  
The thermal limit-cycle instability makes the accretion 
disk jump between the hot stable state and the cool stable 
state.  
When the system has the mass transfer rate ($\dot{M}_{\rm tr}$) 
from the secondary above the critical rate ($\dot{M}_{\rm crit}$), 
it persistently stays in the hot state, which is called 
a nova-like star (NL).  

Z Cam-type DNe are known as an intermediate class between 
DNe and NLs, i.e., $\dot{M}_{\rm tr} \sim \dot{M}_{\rm crit}$, 
in which the light curve alternates between DN outbursts 
and standstills on timescales of a few hundred days.  
The standstill is terminated by fading to the quiescent state 
in Z Cam stars.  
It has turned out that there exist a small number of 
unusual Z Cam stars (called ``anomalous Z Cam Stars'') 
(e.g., \cite{sim11zcamcamp1,szk13iwandv513cas,sim14stchabpcra}) 
in which the standstill is terminated by brightening instead of 
fading.  This brightening is sometimes called `a stunted outburst' 
(e.g., \cite{hon01nloutburst}).  
\citet{kat19iwand} found three more such objects and recognized 
that they exhibit a characteristic light variation: 
a repetition of a quasi-standstill (i.e., a mid-brightness interval 
with (damping) oscillations) terminated by small brightening.  
He also named this class IW And-type DNe.  

Two models have so far been proposed to explain 
the characteristic light variation in IW And-type DNe.  
The first one is proposed by \citet{ham14zcam} who argued 
the variation in the mass-transfer rate explains 
the brightening accompanied by a deep dip in IW And stars.  
The second model is proposed by \citet{kim20tiltdiskmodel} 
who studied the thermal-viscous instability in tilted 
accretion disks.  
The latter authors demonstrated that tilted disks  
can achieve a new kind of accretion cycle as mass supply 
patterns in the tilted disk are quite different from those 
in the usual non-tilted disk.  In fact, the tilted disk is 
inferred in some IW And-type stars, since negative superhumps 
are detected in these stars \citep{arm13aqmenimeri,gie13j1922}.  

Negative superhumps having periods shorter than the orbital 
period and super-orbital modulations having periods of 
a few to several days in CVs are believed to originate 
from the tilted accretion disk (e.g., \cite{bon85tvcol,pat99SH}).  
The periodicity of negative superhumps is now considered to be 
produced by the beat period between the orbital period and 
the period with which the tilted disk retrogradely precesses and 
the light source is most likely variable dissipation 
in the bright spot which is produced by the collision of the gas stream 
with the disk matter, as the gas stream sweeps the disk surface 
with this period \citep{woo00SH,mur02warpeddisk,woo07negSH}.  
According to this interpretation, super-orbital modulations 
are naturally interpreted as the representation of the change 
in the projection area of the tilted disk against observers with 
the precession period.  
The implication of the tilted disk in IW And-type stars 
motivated the authors of \citet{kim20tiltdiskmodel} to study 
the thermal-viscous instability in tilted disks, and hence, 
it becomes more important for us to know its detailed property.  
However, the property of the tilted disk in CVs has not 
so far been well investigated from observations.

A particular system, KIC 9406652, is interesting in 
this respect, because it exhibits typical IW And-type 
light variations and because good evidence of the tilted 
disk was already found in it \citep{gie13j1922}. 
This system was observed by the {\it Kepler} satellite 
during $\sim$4 yrs.  
\citet{gie13j1922} detected three kinds of periodic 
light variations, whose frequencies are 0.242, 3.929, and 
4.171~d$^{-1}$, denoted as $f_1$, $f_2$, and $f_3$, 
by using the {\it Kepler} data.  They also performed 
spectroscopic observations and identified the orbital 
period to be 6.108-hrs which corresponds to $f_2$.  
They confirmed the relation of $f_1 = f_3 - f_2$, and 
hence, argued that $f_1$ and $f_3$ represent the frequency 
of the retrograde precession of the tilted disk and 
the frequency of negative superhumps originating from 
the bright spot on the tilted disk, respectively.  
Besides, they estimated the binary parameters of 
this system.  The mass ratio and the inclination angle 
are determined to be $0.83 \pm 0.07$ and $\sim$50~deg, 
respectively.  Moreover, they estimated how the secondary 
and the disk contribute to the spectral energy distribution 
from ultraviolet to infrared wavelengths.  

In this paper we re-investigate the {\it Kepler} data of 
KIC 9406652 to study the tilted disk in detail, 
just not only for IW And stars but also for CVs in general 
as its {\it Kepler} data have rich information.  
We aim to extend the existing study by \citet{gie13j1922} 
and to extract as much information on the tilted disk 
as possible, by analyzing the time variations of frequencies, 
amplitudes, and light curve profiles of three periodic signals 
in detail.  
By so doing, we also examine the models for IW And-type 
stars so far proposed.  
This paper is structured as follows.  
Section 2 briefly describes the methods of the timing analyses.  
Section 3 presents our analyses of the {\it Kepler} light curves 
and their interpretations.  
In section 4, we discuss our results and examine the two models 
for the IW And-type stars and section 5 is the summary.

\section{Methods of data analyses}

\subsection{Extraction of light curves}

We have extracted the long-cadence {\it Kepler} public 
light curves.  \textcolor{black}{The long-cadence data integrate 
over multiple $\sim$6-s exposures to give 1766-s 
observations and average short-lived events 
\citep{mur12kepler}.}  We use the Simple Aperture 
Photometry (SAP) data.  The count rates are converted to 
a relative magnitude by the formula mag = $14-2.5\log e$ 
where e is the Kepler electron count rate (electrons s$^{-1}$), 
and a constant of 14 is arbitrarily chosen for convenience.  
Hereafter we present all of the observation times in 
Barycentric Julian Date (BJD).

\subsection{Period analyses}

Before performing period analyses, we have subtracted 
the long-term trend of light curves from the observational 
data by locally-weighted polynomial regression 
(LOWESS: \cite{LOWESS}).  
We have to determine the length of the data for which we perform 
the regression and the smoother span influencing the smoothness of 
each data point in this method.  
When subtracting the trend of the overall light curve in 
subsections 3.2 and 3.3, we have performed LOWESS per 2 days with 
a smoother span of 0.4.  
On the other hand, we have divided the light curves into 
quasi-standstills and brightening, and have performed LOWESS for 
each data set to subtract the long-term trend 
in subsection 3.4.  
Then the smoother span ranged between 0.05--0.5 according to 
the light variation of each data set.  

We have used the phase dispersion minimization (PDM) method 
\citep{PDM} for period analyses.  
The 1-$\sigma$ errors of PDM are calculated by the methods 
described in \citet{fer89error} and \citet{Pdot2}.  
A variety of bootstraps are applied to evaluate 
the robustness of the PDM result.  We have prepared 
100 samples, each of which randomly includes half of 
the observations, and performed the PDM analyses 
for the samples. 
The result of the bootstrap is displayed in the form of 
90\% confidence intervals in the resultant statistics.

\textcolor{black}{
In extracting the $O-C$ curve of times of maxima and 
the full-amplitude variation of periodic modulations 
in subsections 3.2 and 3.4, 
we have derived the phase-averaged profile from 
a part of the observational data and have fitted each hump 
by using the profile as the template.  The parameters to be 
obtained in that fitting for each hump are the time of 
the light maximum, from which the $O-C$ is derived, 
the full-amplitude, and the offset of the brightness.  }
Here $O-C$ is defined as the times of maxima minus the constant 
period multiplied by the cycle of each hump and stands for 
the time derivative of periodic variations.  
The details of the method are described in \citet{Pdot}.

\section{Results}

\subsection{Overall light curve}

The overall {\it Kepler} light curve of KIC 9406652 is 
exhibited in the top panel of Figure \ref{ampfreq}.  
This light curve contains additional data of the Kepler 
observation quarter 16 (after BJD 2456392) in comparison 
with the data reported by \citet{gie13j1922}.  
This system shows very clearly the IW And-type phenomenon 
with a cyclic light variation as the quasi-standstill 
\textcolor{black}{is terminated} by brightening.  
Deep dips occasionally occur soon after brightening, 
though one exception was observed around BJD 2455190.  
Hereafter one cycle of the IW And-type phenomenon 
is defined as that from one brightening to the next, 
and we set the start as the beginning of brightening 
and its end as that of the next brightening.  
We display the borders of each cycle as dashed lines 
in the top panel of Figure \ref{ampfreq}.    
The cycle length ranges from 30 to 110~d with 
its average value around 50~d and the amplitudes of 
brightening are less than but close to 1 mag.  
The Kepler data of this system extend over 1500~d and 
we identify 28 cycles as shown in the top panel of 
Figure \ref{ampfreq}. 
We find three types of periodic light variations: 
negative superhumps, orbital signals, and sometimes 
remarkable modulations with a $\sim$4-d period called 
the super-orbital period, as reported by \citet{gie13j1922}.  
We study these three periodic signals in the following 
subsections.

\subsection{Negative superhumps}

\subsubsection{Correlation with the overall light variations}

\begin{figure*}[h]
\begin{center}
\FigureFile(170mm, 50mm){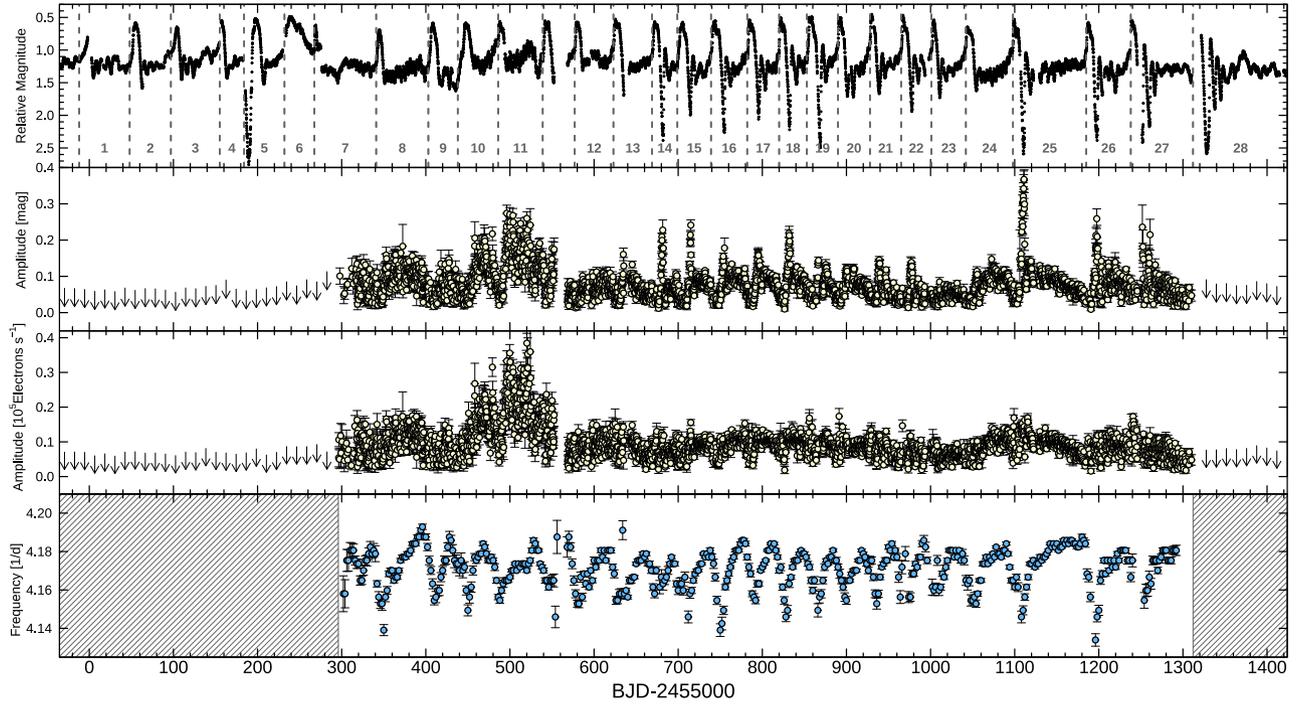}
\end{center}
\caption{Overall light curve, frequency variation, and amplitude variation of negative superhumps in KIC 9406652 during BJD 2454960--2456430.  The first, second, third, and bottom panels represent the overall light curve, which is binned per 0.1~d, the amplitude in the magnitude scale, the amplitude in the flux scale, and the frequency, respectively.  Each cycle of the IW And-type phenomenon is indicated by dashed lines in the top panel.  The arrows in the second and third panels stand for the upper limit of amplitudes.  
}
\label{ampfreq}
\end{figure*}

We first discuss negative superhumps (NSHs) as they are 
the most conspicuous sign of the tilted disk in CVs.   
The second, third, bottom panels of Figure \ref{ampfreq} 
show the time evolution of the amplitude of negative 
superhumps in the magnitude scale, that in the flux scale, 
and of their frequency variation, respectively.    
Negative superhumps were clearly detected around 
BJD 2455300--2456300, and thus the disk is regarded 
to have been tilted in this interval.  
The amplitude of negative superhumps was very small and/or 
the hump shape was unclear before and after this interval, 
so that we were not able to explore the time variation of 
their frequencies and amplitudes.  
The IW And-type phenomenon appeared even when 
negative superhumps were very weak \textcolor{black}{and/or non-existent}.  
This may imply that the IW And-type phenomenon could occur 
regardless of whether the disk is tilted or not.  
The anomalous light variations such as a deep dip around 
BJD 2455190 and long-lasting brightening around 
BJD 2455230--2455290 (cycle 6) were observed just before 
negative superhumps clearly appeared.  
It is not known why this phenomenon occurred and 
\textcolor{black}{whether or not it was related in any way to} 
the development of negative superhumps.

\subsubsection{Frequency variations}

The frequency variation of negative superhumps has 
information about the variation of the radius and 
the radial mass distribution of the tilted disk, 
as discussed in subsection 4.1.  
In this paper, we assume that the disk is rigidly tilted 
and does not have any warped structures.  
The frequency of the nodal precession of the rigidly tilted disk, 
$\nu_{\rm nPR}$, is expressed as 
\citep{pap95tilteddisk,lar98XBprecession}: 
\begin{equation}
\nu_{\rm nPR} = - \frac{3}{8 \pi} \frac{G M_2}{a^3} \frac{\int \Sigma r^3 dr}{\int \Sigma \Omega r^3 dr} \cos \theta, 
\label{nu_nPR}
\end{equation}
where $M_2$ is the mass of the secondary, $r$ is 
the radial distance from the central WD, $a$ is 
the binary separation, $G$ is the gravitational constant, 
$\Sigma$ is the surface density of the disk, $\Omega$ is 
the Keplerian angular velocity of the disk matter, 
and $\theta$ is the tilt angle, respectively.  
The minus sign in this equation means that the nodal 
precession is retrograde.  
The frequency of negative superhumps, $\nu_{\rm NSH}$, 
is given by the synodic frequency between the tilted disk 
and the orbiting secondary star, and it is therefore 
expressed as 
\begin{eqnarray}
\label{nu_nSH}
\nu_{\rm NSH} = \nu_{\rm orb} - \nu_{\rm nPR} &= \nu_{\rm orb} + \frac{3}{8 \pi} \frac{G M_{\rm 2}}{a^3} \frac{\int \Sigma r^3 dr}{\int \Sigma \Omega r^3 dr} \cos \theta \\ 
&\textcolor{black}{= \nu_{\rm orb} + \frac{3}{8 \pi} \frac{\sqrt{G} M_2}{a^3 \sqrt{M_1}} \frac{\int \Sigma r^3 dr}{\int \Sigma r^{3/2} dr} \cos \theta}. \nonumber
\end{eqnarray}
Here $\nu_{\rm orb}$ is the orbital frequency of the binary 
\textcolor{black}{and we use $\Omega \equiv \sqrt{G M_1 / r^3}$,  
where $M_1$ stands for the WD mass.  }
We can safely put $\cos \theta \simeq 1$ in equation (\ref{nu_nSH})
because the tilt angle in KIC 9406652 was found to be very small, 
almost always less than 3 deg in subsection 3.4.3.  
We, therefore, interpret that the frequency variation in 
KIC 9406652 would represent the time variation of the radial mass 
distribution and/or the radius of the tilted disk.  
Here we express the frequencies in two different 
ways in this paper, $\nu$, and $f$, where the former expression 
($\nu$) is that of physical origin and it can take a negative 
value while the latter one ($f$) is the observed frequency of 
a periodic signal and it must be always positive. 
The three frequencies observed by \citet{gie13j1922} correspond 
to $f_1 = | \nu_{\rm nPR} |$, 
$f_2 = \nu_{\rm orb}$, and $f_3 = \nu_{\rm NSH}$, respectively.  

\begin{figure}[htb]
\begin{center}
\FigureFile(80mm, 50mm){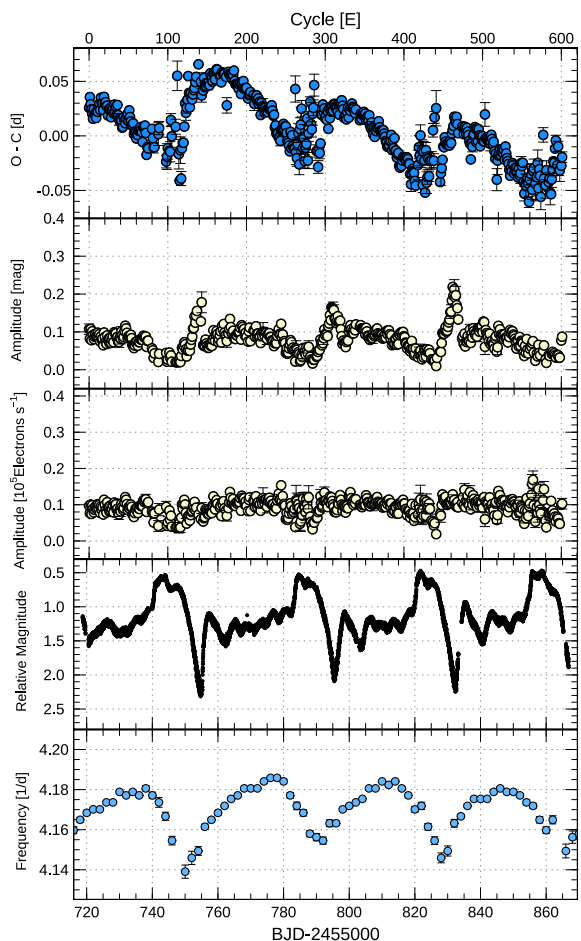}
\vspace{2mm}
\end{center}
\caption{$O-C$ curve, amplitude variation, light curve, and frequency variation of negative superhumps in KIC 9406652 during BJD 2455720--2455865.  Here, we use C = 2455720.7393$+$0.23988 E.  \textcolor{black}{The upper three panels are plotted against the cycle number E and the lower two panels are plotted against BJD, respectively.} }
\label{pick}
\end{figure}

Let us now go back to observations.  
We first subtract the superimposed orbital signal with 
a period of 0.2545~d reported by \citet{gie13j1922} 
to remove resonance signals.  
To extract the frequency variation of negative superhumps, 
we apply PDM in a small window and repeated it by shifting 
the window by a certain time step (see also subsection 2.2).  
The width of the window is 12 d and the time step is 2 d, 
respectively.  
We also divide the interval between BJD 2455296 and BJD 2456312 
into 6 short intervals and derive the $O-C$ curve and the amplitude, 
since the $O-C$ analysis is sensitive to the change of 
the hump shape.  
The resultant times of maxima of negative superhumps are 
given in Tables E1--E6 in the supplementary information.  

The frequency variation takes a regular pattern 
every cycle of IW And-type phenomenon (see the bottom panel of 
Figure \ref{ampfreq}), that is, a sudden decrease in frequency 
by $\sim$0.04-d$^{-1}$ around the initial stage of brightening, 
which is sometimes accompanied with a small and rapid increase, 
and the gradual increase during the quasi-standstill.  
Although the start of decreases seems to sometimes precede 
the rapid rise of the luminosity in Figure \ref{ampfreq}, 
it is simply due to the artifact of our data analyses 
with the 12-d window width as discussed by 
\citet{osa13v344lyrv1504cyg,osa14v1504cygv344lyrpaper3}.\footnote{We 
confirmed that the frequencies jumped in the almost same way 
in analyzing the same data with the 8-d window width.}  
To show the period change more clearly, we pick up 
the time interval during BJD 2455720--2455865 and exhibit 
the results in an expanded form together with the $O-C$ curve 
in Figure \ref{pick}.  
The $O-C$ curve clearly shows the sudden period increase 
at the beginning of brightening, which corresponds to 
the rapid decrease in the frequency.  
The interpretation of the time evolution of the frequency 
of negative superhumps will be discussed in subsection 4.1 
concerning the modeling of the IW And-type phenomenon.

\subsubsection{Amplitude and profile variations}

The amplitude variation in the magnitude scale is displayed in 
the second panels of Figures \ref{ampfreq} and \ref{pick} during 
one cycle of the IW And-type phenomenon.  
The upper limit of amplitudes are estimated by folding 
the 12-d light curves with a period of 0.2401~d before 
BJD 2455296 and with 0.2397~d after BJD 2456312.  
The pattern of amplitude variations in the magnitude scale 
looks just like the inverted one of the light curve, 
except during BJD 2455450--2455550.    
In particular, the spike in the amplitude variation corresponds to 
the luminosity dip in the light curve.  
This means that the {\it absolute} amplitude of negative superhumps 
would stay more or less in constant in time, independent of 
the intrinsic variation of the disk luminosity.  
The amplitude does not change so much in the flux scale as seen 
in the third panels of Figures \ref{ampfreq} and \ref{pick}, 
although it becomes a little smaller during brightening.  
This supports the picture that the source of negative superhumps 
is the change in the luminosity of the bright spot 
as the gas stream sweeps on the tilted disk surface, as proposed 
by \citet{woo00SH} and \citet{woo07negSH}.
It is not known why the amplitude of negative superhumps 
during BJD 2455450--2455550 violently varied both 
in magnitude scale and in the flux scale.  

\begin{figure*}[htb]
\begin{center}
\begin{minipage}{0.49\hsize}
\FigureFile(80mm, 50mm){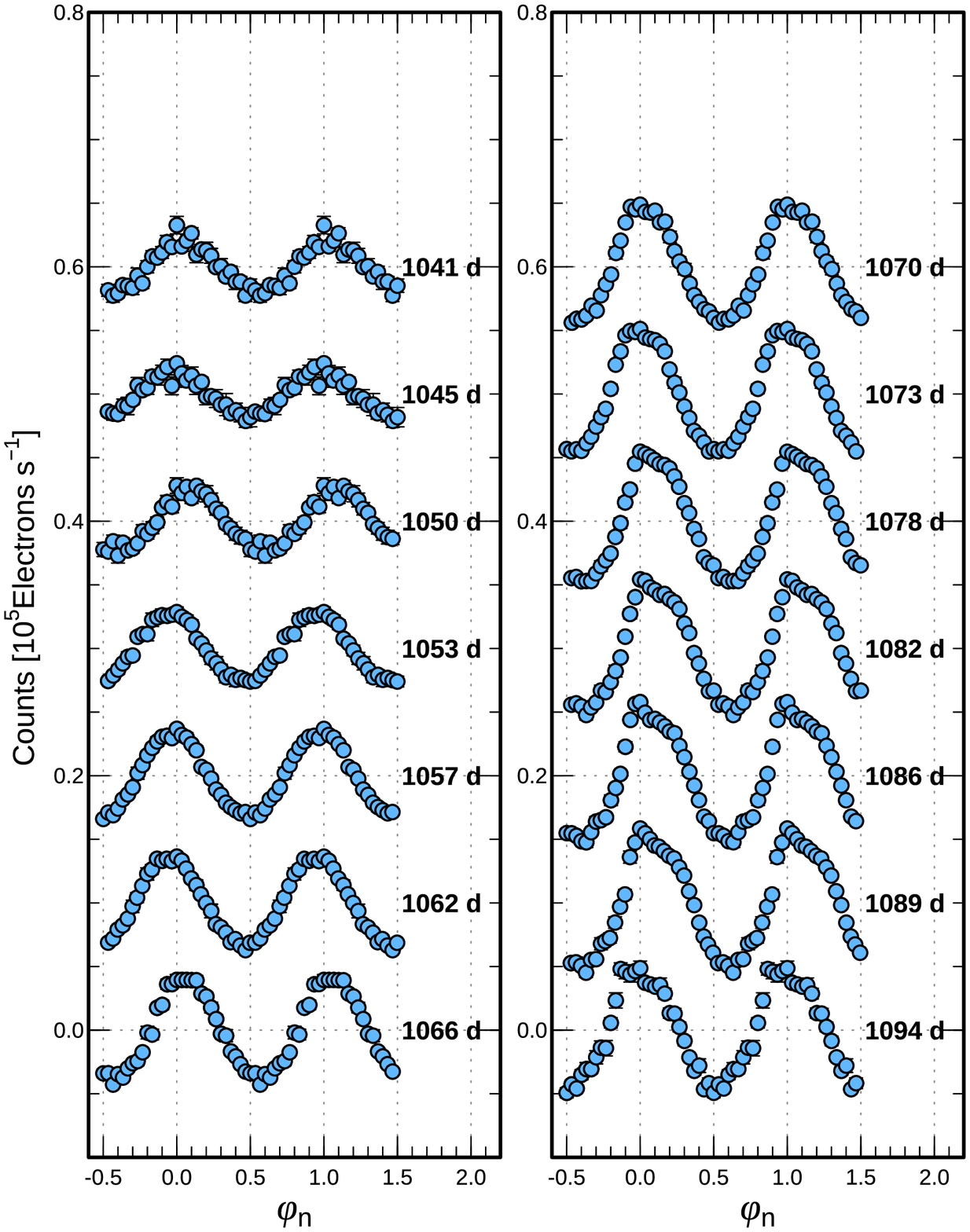}
\end{minipage}
\begin{minipage}{0.49\hsize}
\FigureFile(80mm, 50mm){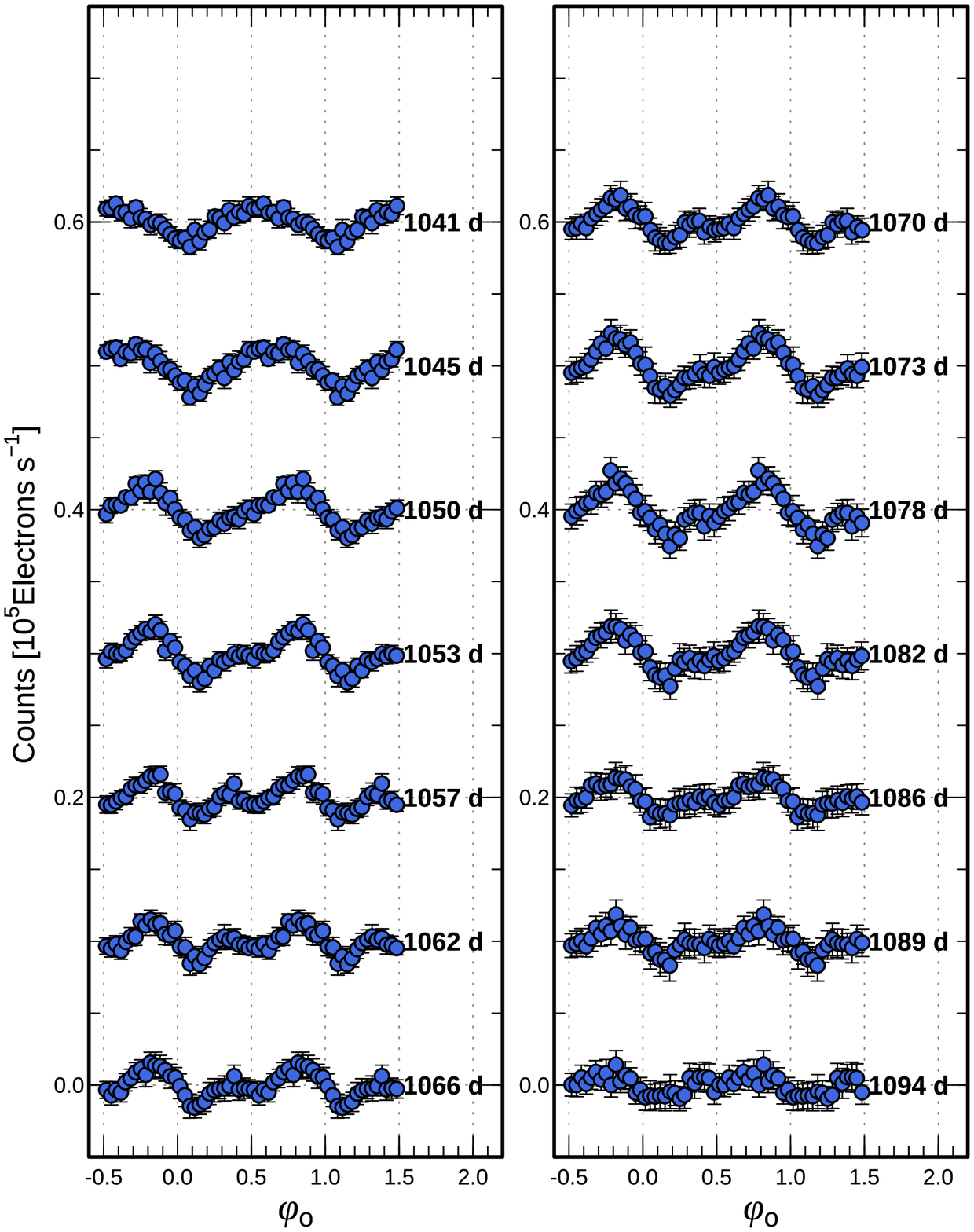}
\end{minipage}
\end{center}
\caption{Time evolution of phase-averaged profiles of negative superhumps and orbital signals during BJD 2456042--2456098 (cycle 24 in Figure \ref{ampfreq}).  In constructing these profiles, we have folded light curves with the superhump period and the orbital period for an interval of 12.7-d (the window) and have repeated it by shifting the window by a 4-d time step.  The number at the right side of each profile represents the date at the center of each window.  To obtain the orbital profile, we have used the epoch of BJD 2456404.971.  Some offsets are added to each profile for visibility in the vertical direction.  }
\label{profile1}
\end{figure*}

We also explore the time evolution in hump profiles.  
The light curve profiles of negative superhumps in 28 cycles 
of IW And-type phenomena are exhibited in the left panels of 
Figures E1--E28 in the supplementary information.  
Here we pick up one panel among them and show it in 
the left panel of Figure \ref{profile1} as an example of 
the light curve profiles.  
We denote the phase of negative superhumps as $\varphi_{\rm n}$ 
to distinguish it from the phase of orbital signal, 
$\varphi_{\rm o}$, and that of super-orbital modulations, 
$\varphi_{\rm s}$.
As discussed above, the light source of the negative superhumps 
is most likely due to variable luminosity of the bright spot of 
the gas stream which sweeps over the face of the tilted disk 
as the secondary star moves around \citep{woo07negSH}.  
As seen in the left panel of Figure \ref{profile1}, 
the \textcolor{black}{light curve} of negative superhump is always single-peaked, 
which would mean that the bright spot formed on the back face 
of the disk is hidden by the optically-thick disk from observers  
as discussed in \citet{woo07negSH}.  
\textcolor{black}{
\citet{woo07negSH} presented their simulated light curves 
in which the variation is approximately single sinusoidal 
with phase (their Fig.~3); however, they did not give 
any explanation for why the light curve is not flat when 
the stream impact point is behind the disk.}  
\citet{kim20tiltdiskmodel} calculated the single-particle 
trajectory of the gas stream from the $L_1$ point 
and determined the position of the disk surface where 
the gas stream hits in the tilted disk. 
In such a picture, the expected hump profile of negative 
superhump would consist of a half sinusoid in one half of 
the period together with the flat-bottom profile 
in the other half, because the bright spot on the backside 
is invisible to observers in the optically-thick disk.
The observational \textcolor{black}{light curve} of negative 
superhumps contradicts with this simple picture as 
the flat-bottom profile was never observed.  

The key to understanding these phenomena would be 
the gas-stream overflow.  
As the gas stream collides with the outer disk rim, 
some part of the stream matter overflows the disk outer edge 
even in the non-tilted disk 
\citep{arm98streamimpact,kun01streamdiskoverflow}, 
and reaches the vicinity of the Lubow-Shu radius \citep{lub75AD}.  
The main part of gas stream collides with the disk outer edge, 
and the remnant part of the stream flows over both above and 
below the disk edge and \textcolor{black}{reaches the inner disk}.  
As shown in subsection 3.4.3, the tilt angle of the disk 
in this system is lower than 3~deg.  
In the slightly tilted disk with such a low tilt angle, 
most of the gas stream may collide with the disk outer edge.  
However, the fraction of the overflowing matter going to 
the upper face of the disk will vary against that going to 
the lower face with the superhump period as the secondary 
orbits around the disk. 
The profile and the amplitude of negative superhumps 
are probably determined by how much of the gas stream 
overflows above the disk edge at each phase of negative 
superhumps, and they depend most likely on the change 
in the thickness of the outer disk rim and/or the tilt angle 
of the disk.  
This model of the gas-stream overflow seems to be consistent 
with a rich variety in the profile such as sinusoidal curves, 
flat-top shapes, triangular waveforms, and so on.  
The small decrease in the amplitude of negative superhumps 
in the flux scale during brightening may originate from the small 
increase in the thickness of the outer disk at the brightening stage, 
as more of the gas stream will be intercepted at the disk edge.

\subsection{Orbital signals}

\begin{figure}[htb]
\begin{center}
\FigureFile(80mm, 50mm){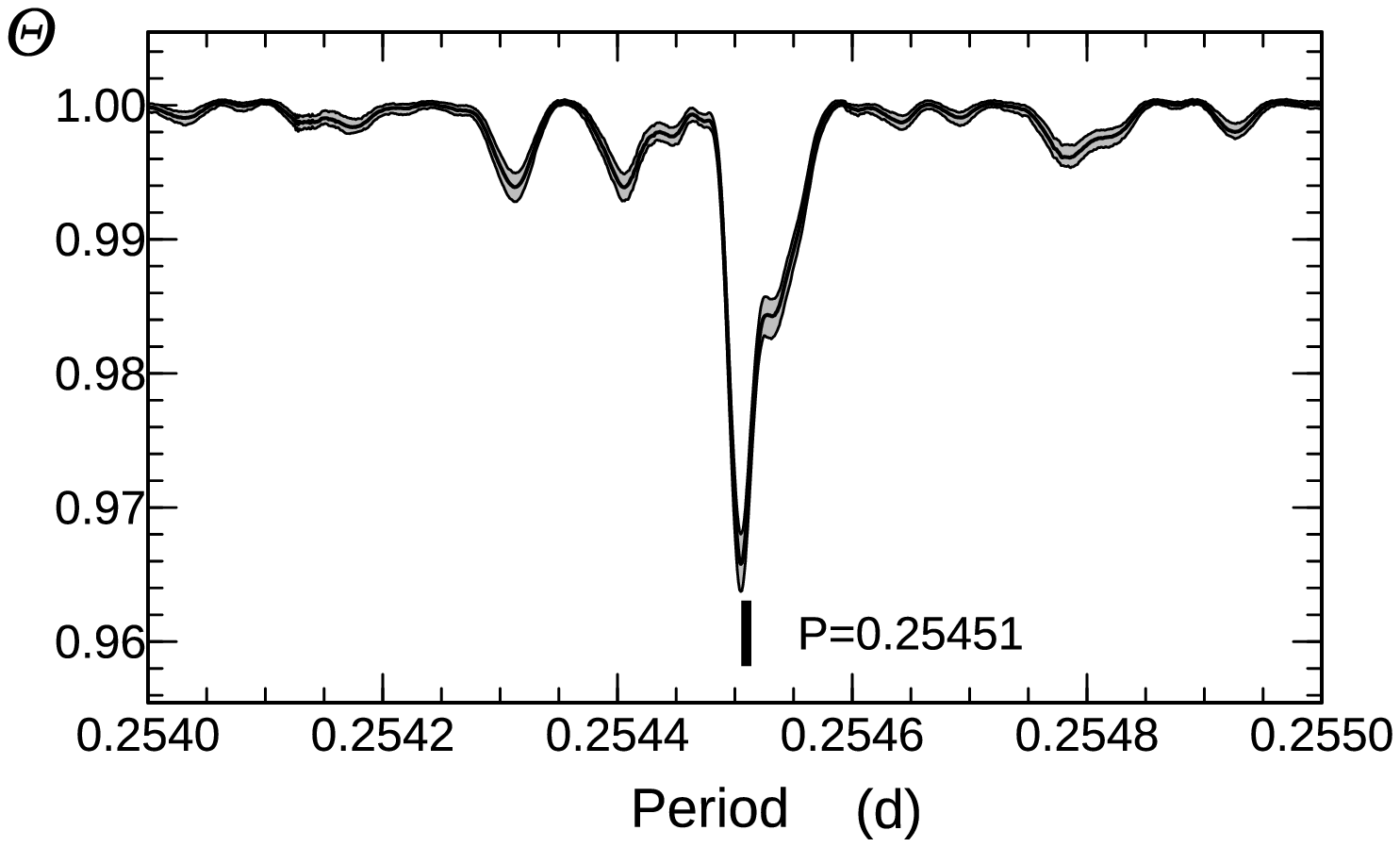}
\\
\vspace{-3mm}
\FigureFile(80mm, 50mm){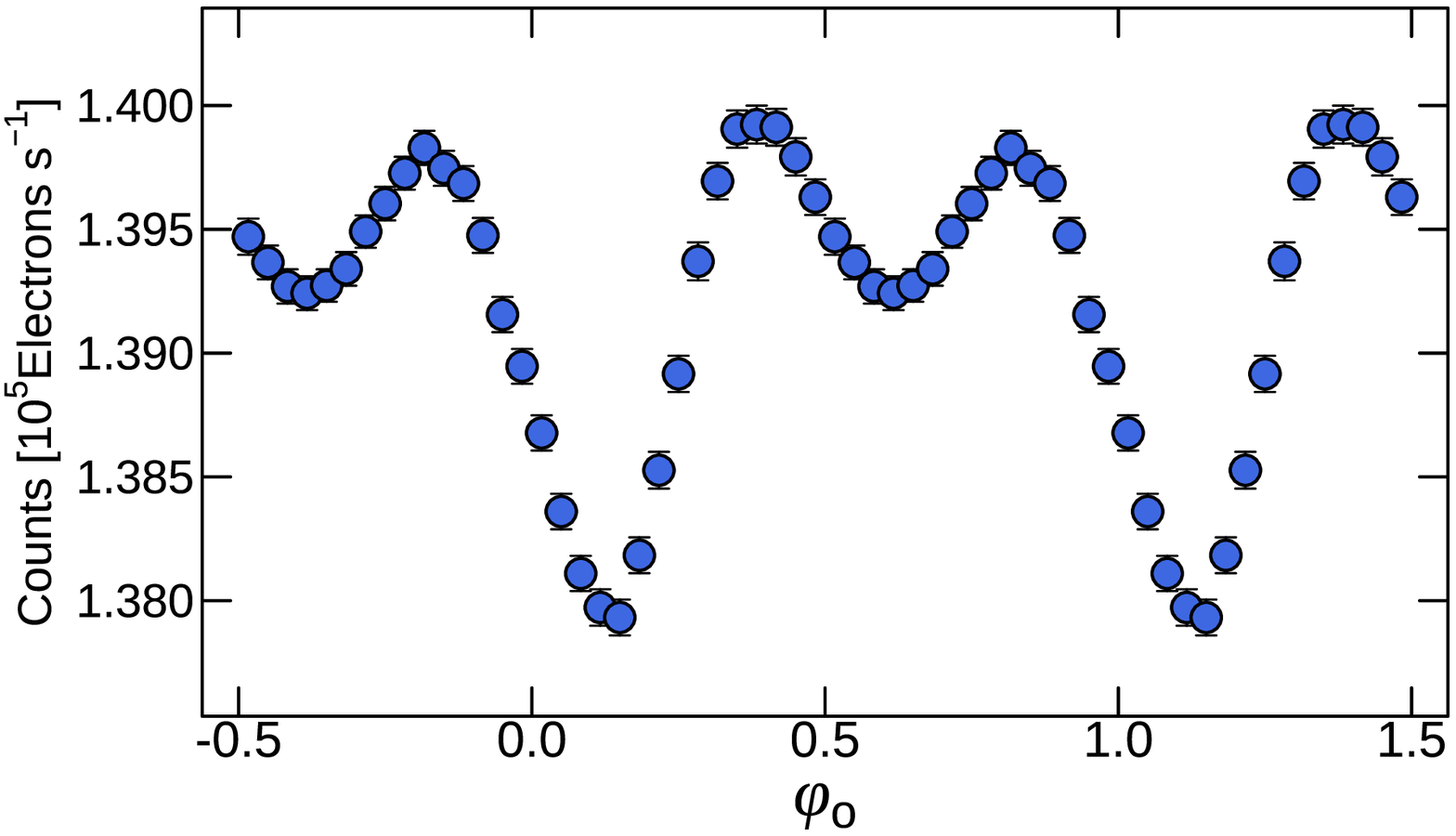}
\end{center}
\caption{Orbital signals in KIC 9406652.  (Upper) $\Theta$-diagram of our PDM analysis.  The area of gray-scale represents 1$\sigma$ errors.  (Lower) Phase-averaged profile.}
\label{orb-pdm}
\end{figure}

In this subsection, we discuss the orbital light curve 
in detail.  To do so, it is the most important to determine 
the orbital period and the binary orbital phase accurately.  
This system, KIC 9406652, is very advantageous in this respect 
because there exists almost continuous Kepler data extending 
over $\sim$1500~d together with the spectroscopic data 
provided by \citet{gie13j1922}.  
The spectroscopic data were obtained within the period 
overlapping with the Kepler data, thanks to \citet{gie13j1922}. 
We have first estimated the orbital period to be 0.2545094(7)~d 
from all of the Kepler data by PDM.  The PDM result is 
displayed in the upper panel of Figure \ref{orb-pdm}.  
We have also estimated the epoch of the inferior conjunction 
of the secondary star to be BJD 2456404.971 by fitting 
the radial velocity (RV) curve by a sinusoidal curve based on 
the data given in Table 2 of \citet{gie13j1922}.  
Hereafter we define the epoch of the inferior conjunction 
of the secondary to be the orbital phase 0.0.  
The lower panel of Figure \ref{orb-pdm} shows  
the phase-averaged profile of the light curves folded 
with the estimated orbital period and epoch.  
The accumulated error in the orbital phase over $\sim$1500~d 
is estimated to be $\sim$0.06, which is less than 0.1, 
when considering the 95\% confidence interval of the estimate 
of the orbital period, so that we can discuss the orbital phase 
with confidence for the data extending over $\sim$1500~d.  

\begin{figure}[htb]
\begin{center}
\FigureFile(80mm, 50mm){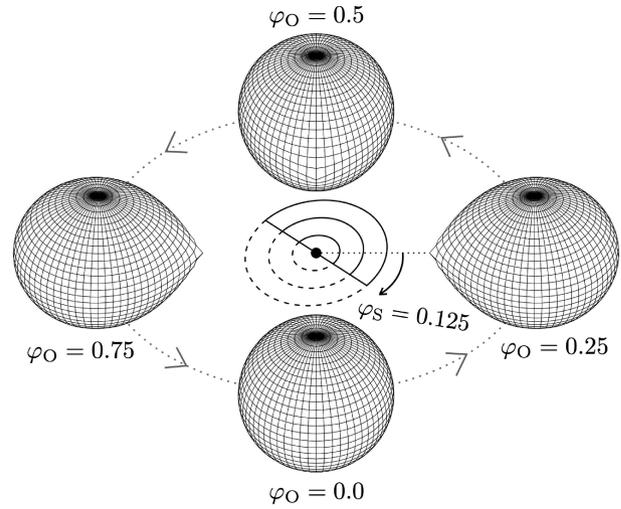}
\end{center}
\caption{Schematic figure of the orbital motion of the secondary star and the retrogradely precessing tilted disk in KIC 9406652.  The inclination angle is assumed to be 50 deg.  The central black point represents a WD and we put a tilted disk at the center with the tilt angle of 10 deg for visibility.  The nodal line is indicated by a diametric line passing the central WD.  The secondary orbits counterclockwise around the central WD and the tilted disk precesses clockwise.  The solid and dashed lines represent the concentric rings of the tilted disk above and below the orbital plane, respectively.  The phases, $\varphi_{\rm s}$ and $\varphi_{\rm o}$, stand for that of the precession of the tilted disk (or that of the super-orbital modulation) and the orbital phase, respectively.  The phase zero for $\varphi_{\rm s}$ is defined when the nodal line of the tilted disk is perpendicular to the line of sight and the front face of the tilted disk turns towards the observers.  The phase of the tilted disk shown in this figure is 0.125.  The orbital phase zero is defined when the secondary star is at the inferior conjunction.  
}
\label{ellip-config}
\end{figure}

As demonstrated below, the light sources of the orbital 
signal consist of three components, that is, 
(1) the orbital hump due to the bright spot formed 
at the outer disk rim, 
(2) the irradiation of the secondary star by the tilted disk 
and/or the WD, 
(3) the ellipsoidal modulation of the secondary star.  
The orbital hump due to the bright spot at the disk rim is 
the most conspicuous feature in the optical light curve of 
CVs with high orbital inclination (see e.g., \cite{krz65ugem}).  
It is caused when the bright spot is on the side of the disk 
facing the observer, peaking at the orbital phase 
$\varphi_{\rm o}=$~0.8--0.9, and it takes a half sinusoidal 
waveform.  Hereafter $\varphi_{\rm o}$ signifies the orbital 
phase.  
If the disk is tilted, the surface of the secondary will 
be more irradiated by the accretion disk than the otherwise 
case, because the inner hot part of the disk would be easily 
exposed to the surface of the secondary.  
The flux variation by the source (2) is predicted to peak 
around $\varphi_{\rm o} = 0.5$ since the irradiated hemisphere 
of the secondary is directly facing the observer at 
the superior conjunction of the secondary.  
The flux variation by the source (3) peaks at 0.25 and 0.75 
in the orbital phase.  We give a schematic picture of 
the orbital motion of the secondary star and the tilted disk 
in Figure \ref{ellip-config} for the binary parameters 
corresponding to KIC 9406652, i.e., a Roche-lobe filling 
semi-detached binary with the mass ratio $q = M_2/M_1 = 0.83$ 
and the orbital inclination, $i = 50$ deg.  

\begin{figure}[htb]
\begin{center}
\FigureFile(80mm, 50mm){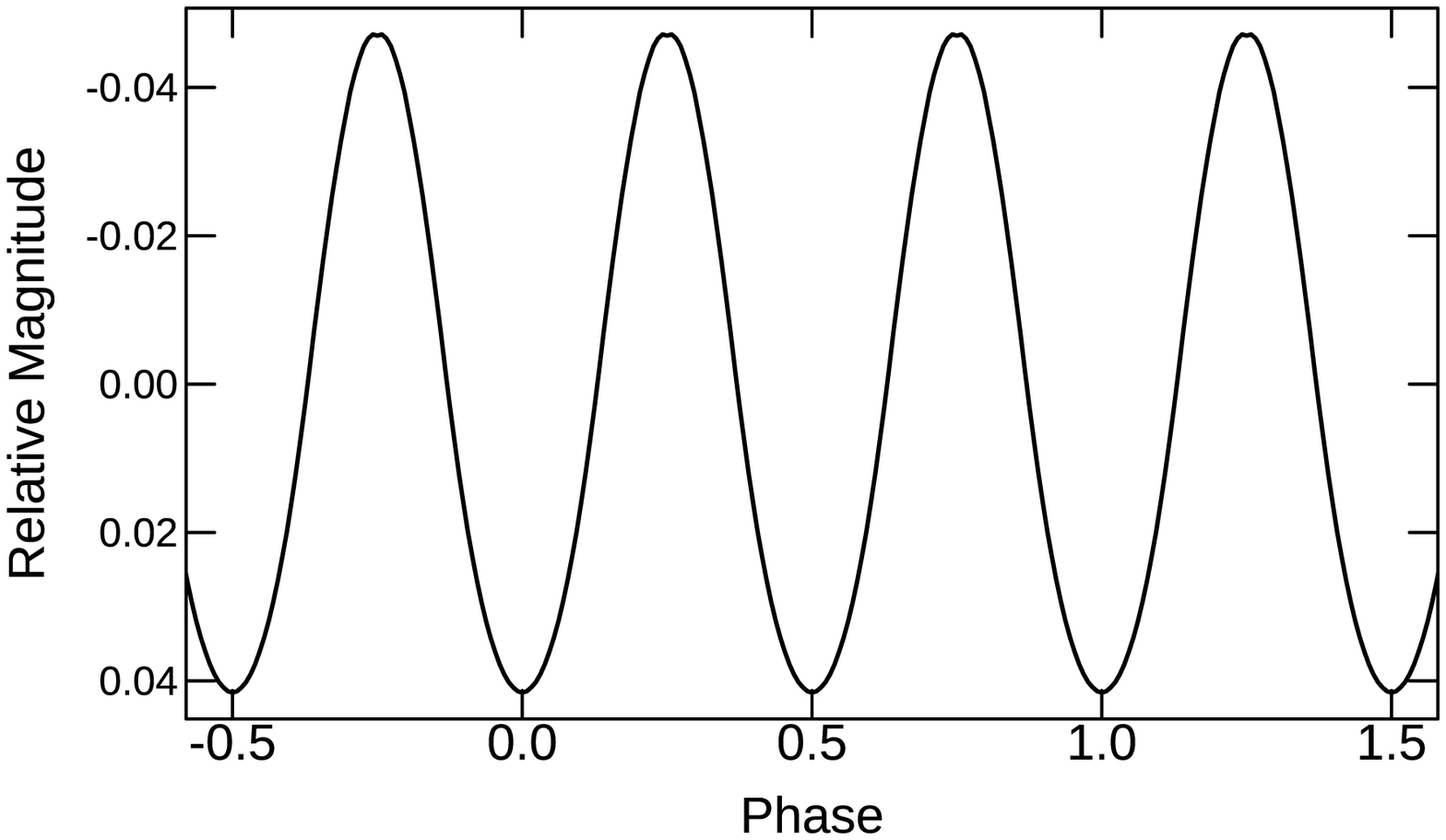}
\\
\vspace{-3mm}
\FigureFile(80mm, 50mm){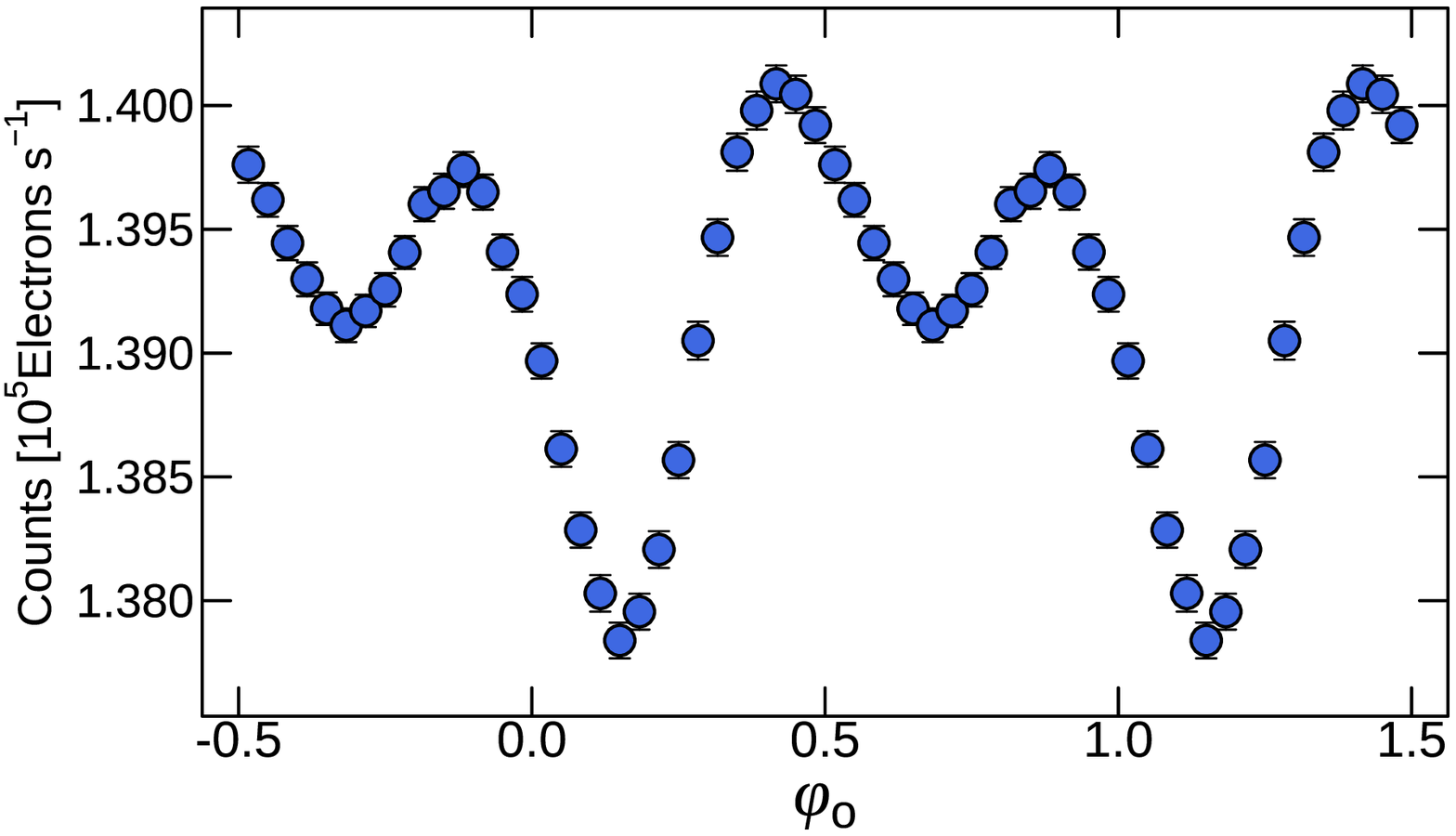}
\end{center}
\caption{(Upper) Model light curve of ellipsoidal modulations of the secondary having 4500~K in the system with the inclination of 50~deg.  (Lower) Phase-averaged profile of overall orbital signals after the subtraction of the ellipsoidal effect.  
}
\label{ellip}
\end{figure}

We first consider the effect of ellipsoidal modulations.  
We can approximately estimate this effect as follows.  
The ellipsoidal variation of a Roche-lobe filling secondary 
star is calculated by using the numerical code described 
in detail by \citet{hac01RN}.  
As for the input parameters in this code, we adopt that 
the temperature of the secondary, the inclination of 
the binary system, the WD mass, and the binary mass ratio 
are 4500~K, 50~deg, 0.9~$M_{\solar}$, and 0.83, respectively, 
according to \citet{gie13j1922}.  
The resultant ellipsoidal modulation is displayed in 
the upper panel of Figure \ref{ellip}.  
We use the following equation to estimate the flux 
variation by the ellipsoidal effect.  
\begin{equation}
F_{\rm ellip} = F_{\rm mean} \frac{F_2}{F_{\rm disk} + F_2} (10^{\frac{\Delta m}{-2.5}} - 1).
\label{ellip-eq}
\end{equation}
Here $F_{\rm mean}$, $F_2$, $F_{\rm disk}$, and $\Delta m$ 
are the mean observational flux, the flux of the secondary, 
the flux of the disk, and the variation of the model light curve 
for the ellipsoidal effect in the magnitude scale, respectively.  
To evaluate $F_2 / (F_{\rm disk} + F_2)$, we use 
the spectral energy distribution given in Figure 10 of 
\citet{gie13j1922}.  
These authors reported that $F_2 / F_{\rm disk}$ is 0.04(1) 
at 5400\AA.  
The value of $F_2 / F_{\rm disk}$ through the {\it Kepler} 
response function is then calculated to be 0.06(2).  
We adopt the average value, 0.06, as 
$F_2 / F_{\rm disk}$.\footnote{Actually, $F_2 / F_{\rm disk}$ 
changes with time, but we neglect its time variation.}  

We can now subtract the ellipsoidal effect from 
the orbital profile by using equation (\ref{ellip-eq}).  
After subtracting the estimated ellipsoidal effect, 
the orbital \textcolor{black}{light curve} given in the lower panel of 
Figure \ref{orb-pdm} is then modified to that shown 
in the lower panel of Figure \ref{ellip}.  
The ellipsoidal effect is only a few percent of 
the entire orbital variation and the major part of 
the orbital signal comes from the orbital hump and 
the irradiation effect.  
Nevertheless, we can confirm that the two peaks around 
0.4--0.5 and 0.8--0.9 become clearer after the subtraction. 
In particular, the peak around 0.4--0.5 becomes higher, 
and the peak around 0.8--0.9 slightly shifts 
towards the orbital phase 1.0.  As expected, the former is 
due to the irradiation of the secondary star and 
the latter is due to the bright spot at the disk rim.  

Since no light variation of the secondary star itself 
is expected to occur, it is thought that the light 
variation due to the ellipsoidal effect is not 
variable in time. 
On the other hand, the light variations due to the other 
two sources, (1) the orbital hump by the bright spot and 
(2) the irradiation of the secondary, vary with time because 
they depend on the disk geometry and the disk luminosity.
By utilizing their time variations, we can isolate 
one component from the other.  
We have investigated the time evolution of the orbital light 
profiles in almost the same way as we have done for negative 
superhumps and display them in the right-hand 
panels of Fig.~E1--E28 in the supplementary information 
for 28 cycles of the IW And-type phenomenon.  
The width of the window used is 
$50 \times P_{\rm orb} \simeq 12.7$~d, 
which is the same as that in the case of negative 
superhumps.  

\begin{figure}[htb]
\begin{center}
\FigureFile(80mm, 50mm){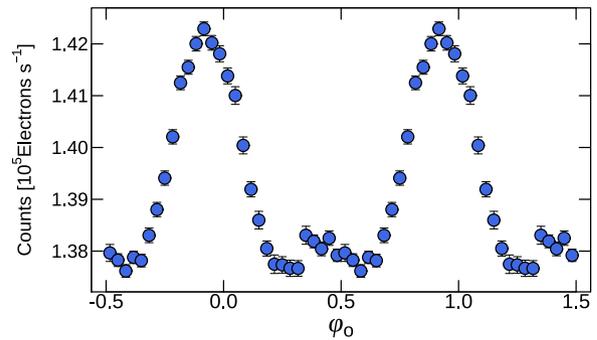}
\end{center}
\caption{Phase-averaged profile of orbital signals during BJD 2454964--2455020.  }
\label{orb-initial}
\end{figure}

It is expected that the orbital hump by the bright spot 
will dominate when the disk is not tilted.  
We have confirmed this as light curves were  
those with a single peak at $\varphi_{\rm o} = 0.8-0.9$
when negative superhumps were almost invisible or very weak 
in the earliest phase of the light curve in the top 
panel of Figure \ref{ampfreq} (see also Fig.~E1 and E5).  
Figure \ref{orb-initial} shows the orbital profile derived 
from the data during BJD 2454964--2455020, which is 
the earliest part of the observational data.  
This \textcolor{black}{light curve profile} is close to the half sinusoidal wave, 
a typical orbital hump light curve due to the bright spot.  

\begin{figure}[htb]
\begin{center}
\FigureFile(80mm, 50mm){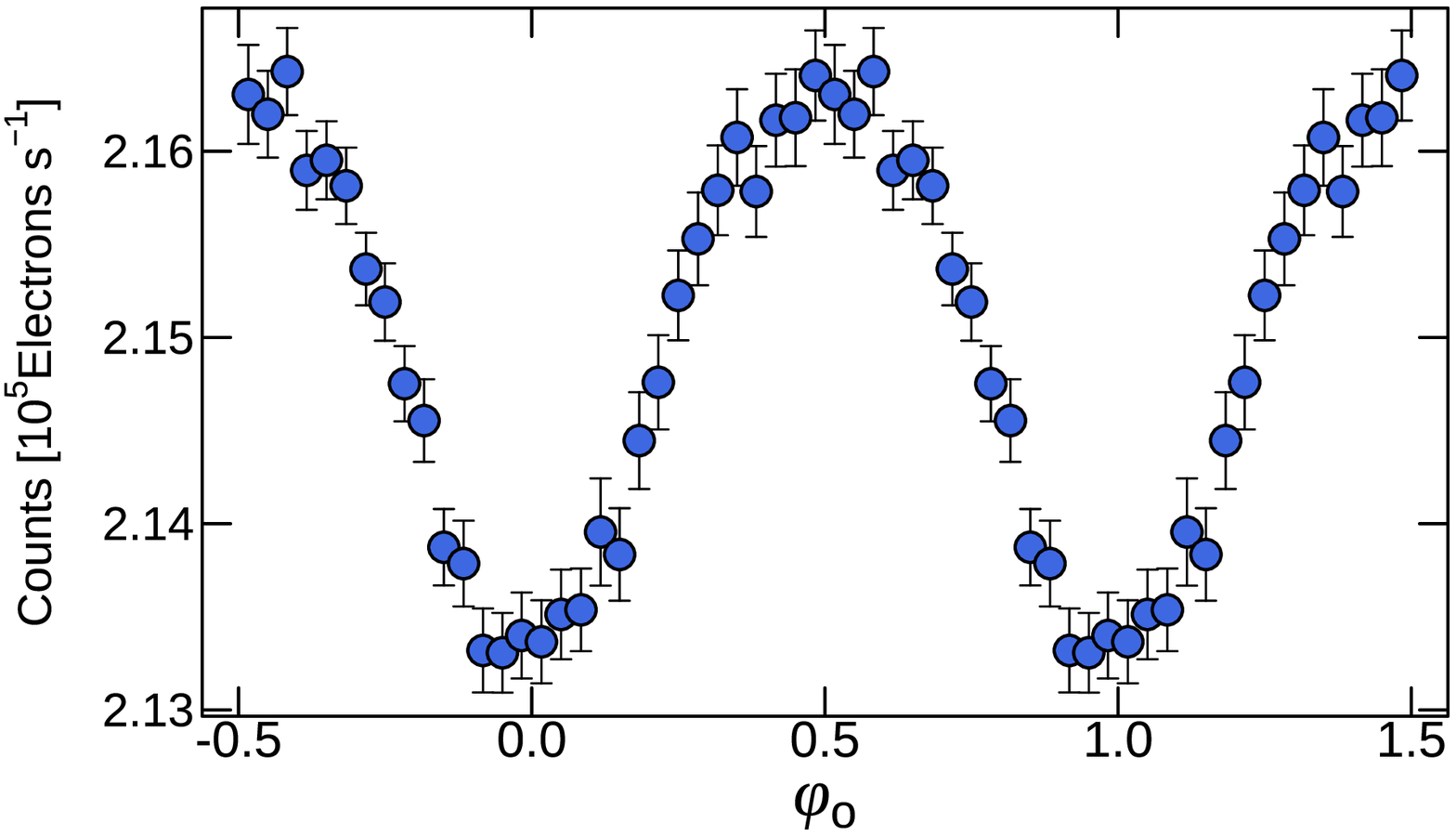}
\\
\FigureFile(80mm, 50mm){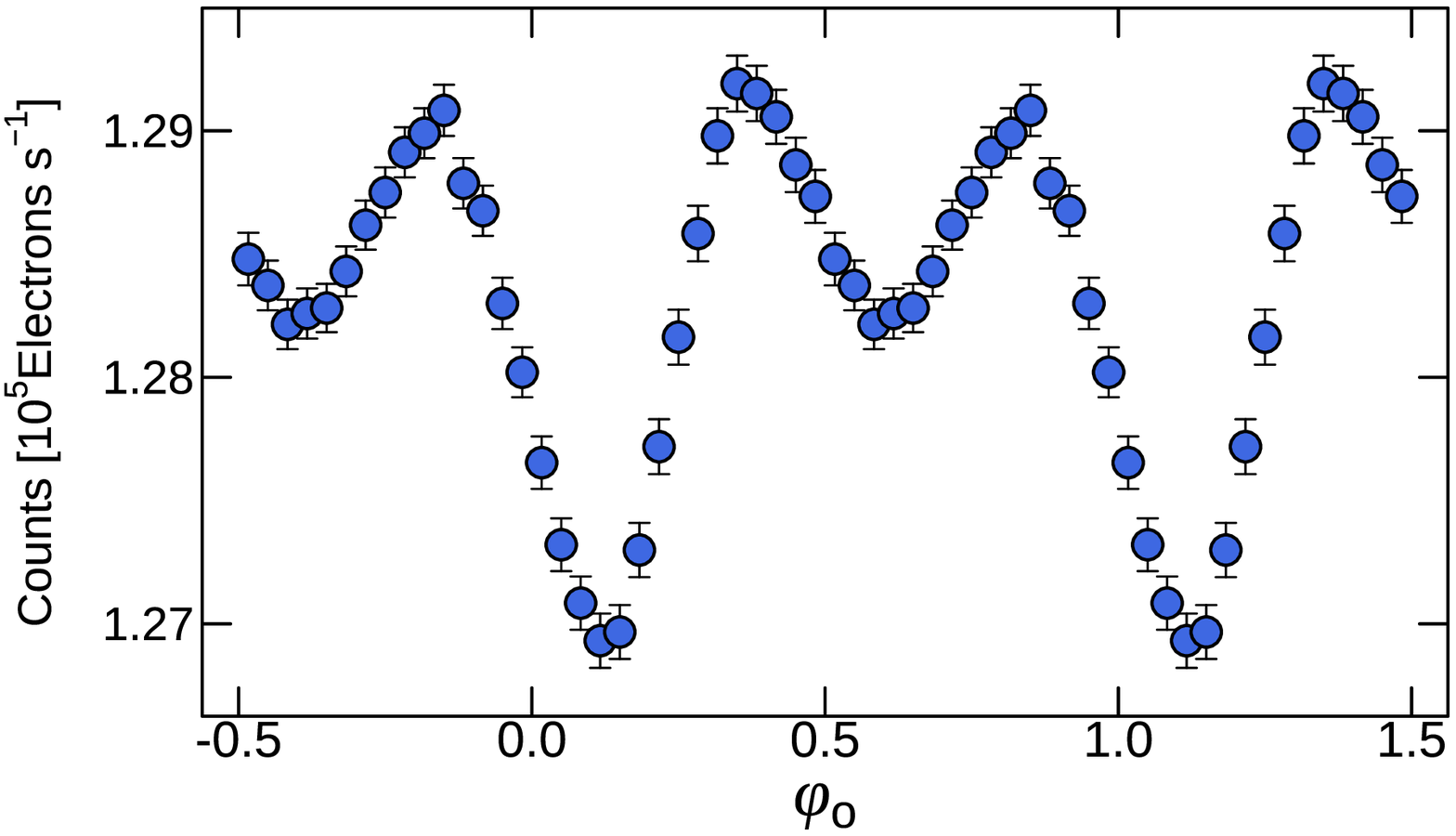}
\\
\FigureFile(80mm, 50mm){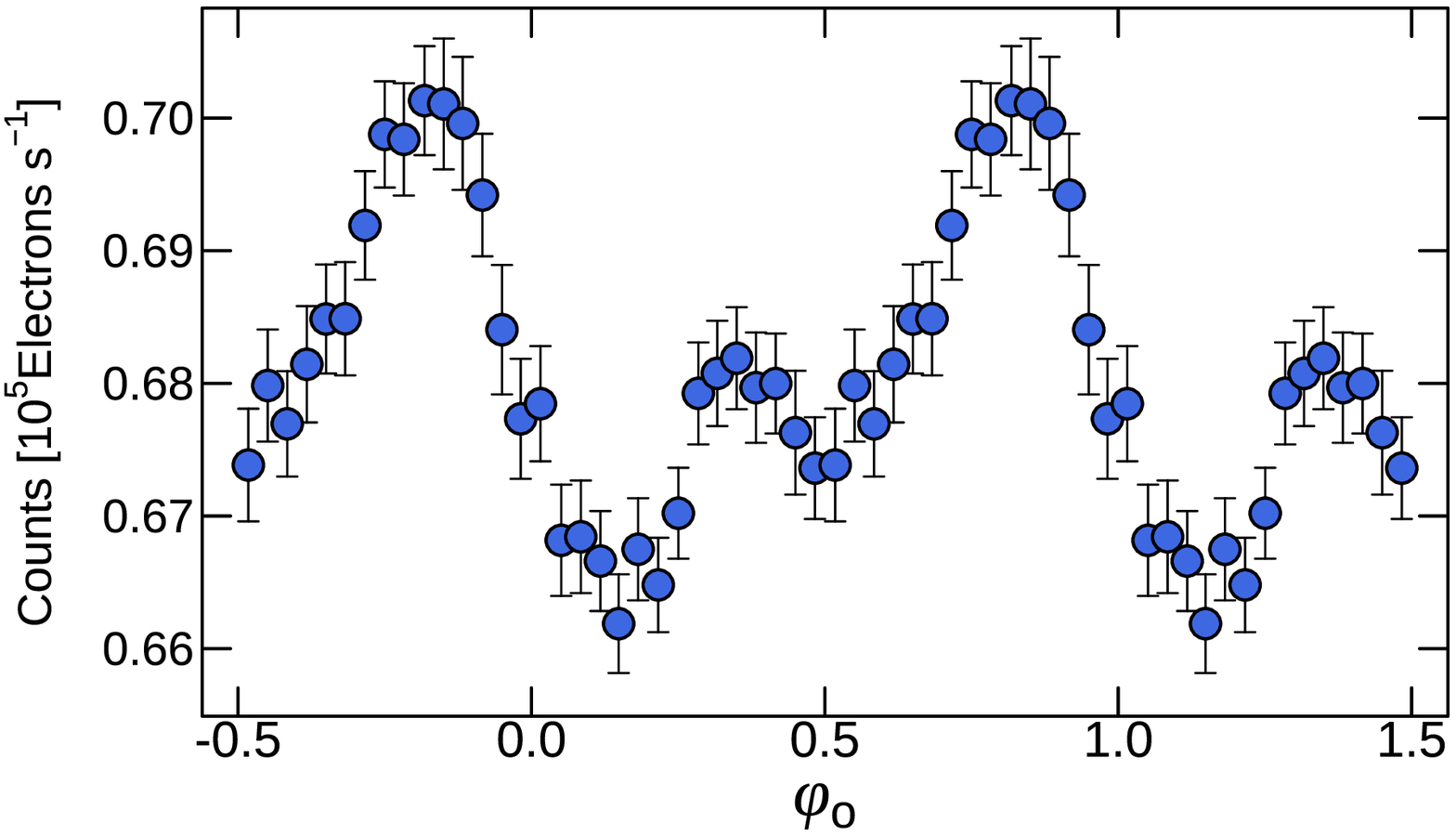}
\end{center}
\caption{Phase-averaged profiles of orbital signals during BJD 2455296--2456312.  (Upper) Profile during brightening.  (Middle) Profile during quasi-standstills.  (Lower) Profile during dips.}
\label{orb2}
\end{figure}

On the other hand, when negative superhumps are clearly 
visible and so the disk is thought to be tilted, the orbital 
light curve becomes very complex as both the bright spot and 
the irradiation of the secondary contribute to them  
(see Fig.~E2--E4, E7--E9, and E12--E27).  
We give an example of the profile variation during 
one cycle (cycle 24) of the IW And-type phenomenon 
in the right panel of Figure \ref{profile1}.  
We find that the peak around the orbital phase 
0.4--0.5, which is thought to originate from 
the irradiation, becomes dominant when the system brightens 
(at the brightening phase of the IW And-type light variation).  
Figure \ref{orb2} exhibits the phase-averaged profiles 
at brightening, quasi-standstills, and dips for about 1000~d 
(BJD 2455296--2456312), over which sustained negative 
superhumps were confirmed.  
The data of brightening, quasi-standstills, and dips have 
count rates more than $1.8 \times 10^5$, between $0.9 \times 10^5$ 
and $1.8 \times 10^5$, less than $0.9 \times 10^5$, respectively.  
We see that the irradiation effect appearing at 
$\varphi_{\rm o} = 0.5$ becomes more dominant 
when the system becomes brighter, 
while the peak around phase 0.8--0.9 becomes remarkable 
when the system becomes fainter.  
The major part of the profile at the brightening stage 
would originate from the irradiation effect, and 
the orbital hump is the strongest at dips.  
The profile at quasi-standstills 
is the intermediate one between the other two.  
These observations are consistent with our expectation that 
the secondary star will more easily be irradiated by 
the WD and the disk when the disk is tilted, and  
that more flux will illuminate the secondary when the disk 
becomes brighter.  
The light maximum seems to be shifted slightly from phase 0.5 
to the early phase especially during quasi-standstills 
(see the lower panel of Figure \ref{orb2}).  This could be 
because the irradiated region on the surface of the secondary 
may be asymmetric to the line connecting the WD and 
the secondary as a part of irradiating radiation from the disk 
and the WD may be intercepted by the gas stream itself or 
a flared disk rim where the gas stream first collides.  

\begin{figure}[htb]
\begin{center}
\FigureFile(80mm, 50mm){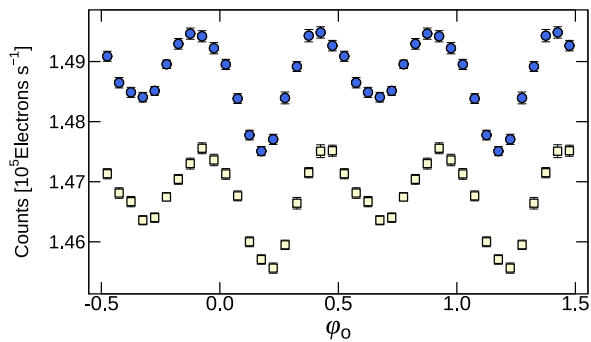}
\end{center}
\caption{Phase-averaged profile during the {\it Kepler} quarters 1--4.  The upper \textcolor{black}{light curve profile} (shown in dots) is that constructed based on the period and the epoch estimated in this paper and the lower one (shown in squares) is that based on the period and the epoch estimated in \citet{gie13j1922}, where the latter is shifted by $- 0.02 \times 10^{5}$~e$^{-}$~s$^{-1}$ from the former.  }
\label{test-orb}
\end{figure}

\citet{gie13j1922} have constructed orbital light curve 
from {\it Kepler} observation quarters 1 to 4 and 
they have noticed that it resembles that of a low-amplitude 
ellipsoidal binary light curve with two unequal minima.  
However, they have chosen the epoch phase arbitrarily.  
Since we have determined the orbital period and the epoch of 
this system quite accurately, we can study the orbital 
light curve for the same interval as that of \citet{gie13j1922}.  
Figure \ref{test-orb} illustrates the orbital light curve 
for the same interval constructed based on the period, 
0.2545094 d, and the epoch, BJD 2456404.941, 
as the phase zero for the inferior conjunction of 
the secondary star in our analysis, together with 
that based on the orbital period and the epoch reported 
by \citet{gie13j1922}\footnote{As for the epoch, 
we averaged the 4 values of $T$ in Table 3 in 
\citet{gie13j1922}}. 
We see that the difference in the phase of the two light curves 
is less than 0.05, which is reasonable as the orbital period 
reported by \citet{gie13j1922} is consistent with that of 
our estimate within the 95\% confidence level.  
This indicates that the orbital phase determined 
in our analysis may be reliable over $\sim$1500~d.  
Although they have stated that it resembles that of 
an ellipsoidal binary light curve, we see from 
Figure \ref{test-orb} that the phases of the double peaks 
do not match those of the ellipsoidal modulation.  
In addition, the two peaks did not necessarily appear 
simultaneously.  One peak appeared sometime in one orbital 
cycle and the other peak appeared in other time 
(see Fig.~E1--E5), and we find clear double peaks 
only after folding light curves for all cycles.  We thus 
conclude that the apparent double peaks are very unlikely 
to be due to the ellipsoidal modulation of the secondary 
but rather due to the superposition of the irradiation of 
the secondary and the orbital hump.  

One might ask why the bright spot was still 
visible in the orbital light curve when the disk was apparently 
tilted, as negative superhumps and orbital signals 
are more or less mutually exclusive to each other in other 
stars (see, e.g., \cite{osa13v344lyrv1504cyg}). 
This is probably because the tilt angle was rather low 
as shown in subsection 3.4.3 and the orbital inclination was 
relatively high in KIC 9406652.  
When the disk is slightly tilted, some part of the gas stream 
would overflow the outer disk edge, while some other part of 
the gas stream would collide with the disk rim as discussed 
in the previous subsection.  
The bright spot is thus formed at the disk rim in this case 
even if the disk is tilted and it is observable as the orbital 
hump if the inclination is sufficiently high.  
The objects reported by \citet{osa13v344lyrv1504cyg} would have 
small inclination angles and/or highly tilted disks.

\subsection{Super-orbital modulations}

\subsubsection{Frequency variations}

\begin{figure*}[htb]
\begin{center}
\FigureFile(160mm, 50mm){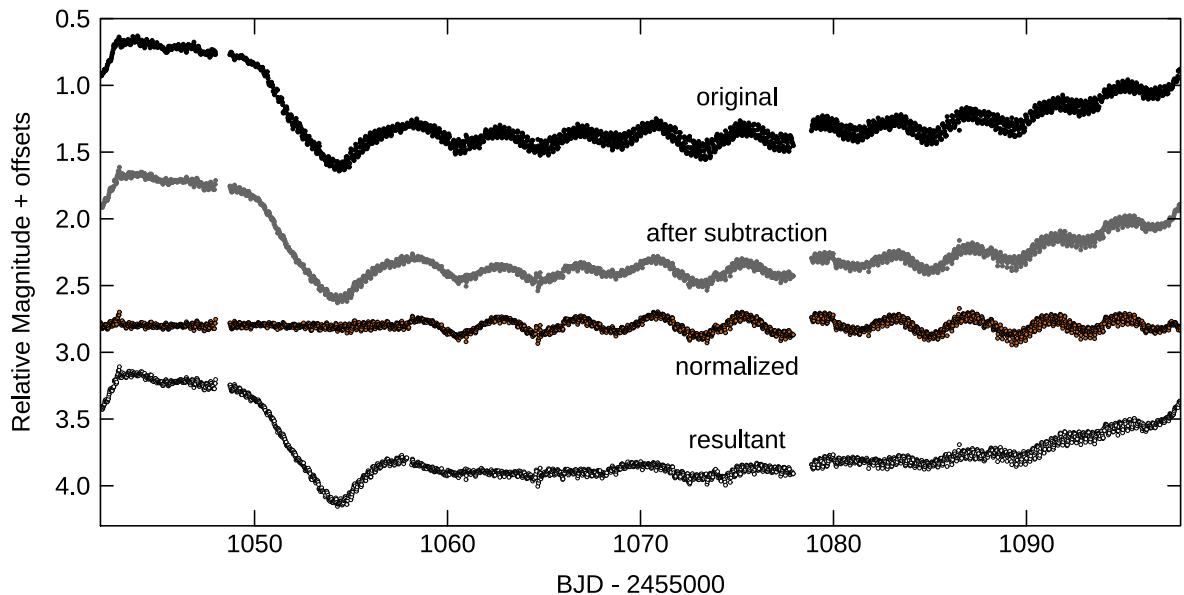}
\end{center}
\caption{Example of the process of the subtraction of the effects of negative superhumps, orbital signals, and super-orbital modulations from the original light curve in the case of cycle 24.  The top light curve (shown by the black dots) is the original light curve and the second one (the gray dots) is that where the effects of negative superhumps and orbital signals are subtracted (shifted by 1 mag for visibility).  The third one (the orange dots) is the normalized light curve where the long-term trend is subtracted from the second one (shifted by 2.8 mag). The bottom one (the open dots) is the resultant light curve where the all effects (superhumps, orbital signals, and super-orbital modulations) are subtracted from the original light curve (shifted by 2.5 mag).  }
\label{24cycle}
\end{figure*}

This object exhibited super-orbital light modulations 
with a $\sim$4-d period besides negative superhumps 
when negative superhumps appeared \citep{gie13j1922}.  
We first discuss frequency variations of super-orbital 
modulations. 
The super-orbital light modulation is thought to be 
produced by variations in the projected area of the disk 
to the line of sight as the nodal line of the tilted disk 
precesses retrogradely. 
On the other hand, negative superhumps are produced by 
variations in the luminosity of the bright spot, which is 
in turn produced by the gas stream sweeping on the tilted disk 
surface as described in subsection 3.2.3.
The period of negative superhumps is given by the synodic 
period between the retrogradely precessing tilted disk and 
the orbiting secondary star.  
Although these two phenomena are observationally completely 
independent, their frequencies are related to each other, i.e., 
by the relation $f_{\rm sup} = f_{\rm NSH} - f_{\rm orb}$, 
where $f_{\rm sup}$, $f_{\rm NSH}$, and $f_{\rm orb}$ denote 
the frequencies of super-orbital modulations, negative 
superhumps, and orbital signals, respectively, 
and correspond to $f_1$, $f_3$, and $f_2$ in \citet{gie13j1922}.  
In addition, $f_{\rm sup}$ stands for $| \nu_{\rm nPR} |$ 
as defined in subsection 3.2.2 (see also equation (\ref{nu_nPR})). 
This means that we can examine the variation of nodal 
precession rates by two different methods, i.e., 
the direct observation of the period of super-orbital 
modulations and the indirect way from the frequency of 
negative superhumps.  

\begin{figure*}[htb]
\begin{center}
\FigureFile(160mm, 50mm){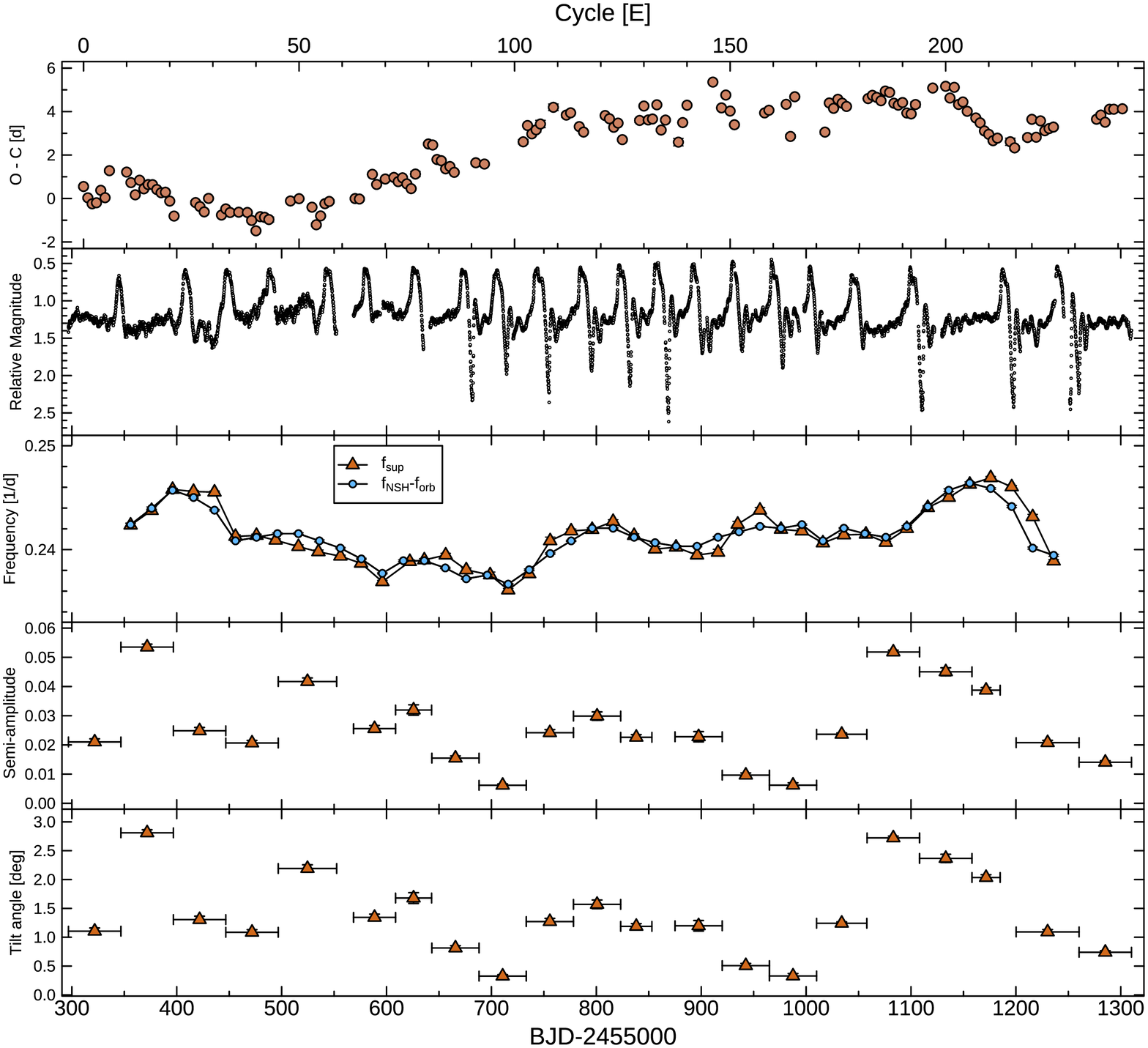}
\end{center}
\caption{Results of analyses of super-orbital modulations in KIC 9406652.  \textcolor{black}{The top panel is plotted against the cycle number E and the other panels are plotted against BJD, respectively.}  (Top) $O-C$ curve of the light maxima.  The assumed period is 4.128~d, and the epoch is BJD 2455298.275, respectively.  (Second) Predicted light variations \textcolor{black}{originating from the accretion process}.  We derived them by subtracting super-orbital modulations, negative superhumps, and orbital signals and binning them per 0.1~d.  (Third) Frequency variations.  The triangles and circles stand for the frequency of super-orbital modulations ($f_{\rm sup}$) and the frequency of negative superhumps minus the orbital frequency ($f_{\rm NSH} - f_{\rm orb}$), respectively.  (Fourth) Semi-amplitude variations \textcolor{black}{in units of relative flux}.  (Bottom) Time evolution of tilt angles of the disk estimated from the semi-amplitude variation.  
}
\label{4days}
\end{figure*}

\begin{figure}[htb]
\begin{center}
\FigureFile(80mm, 50mm){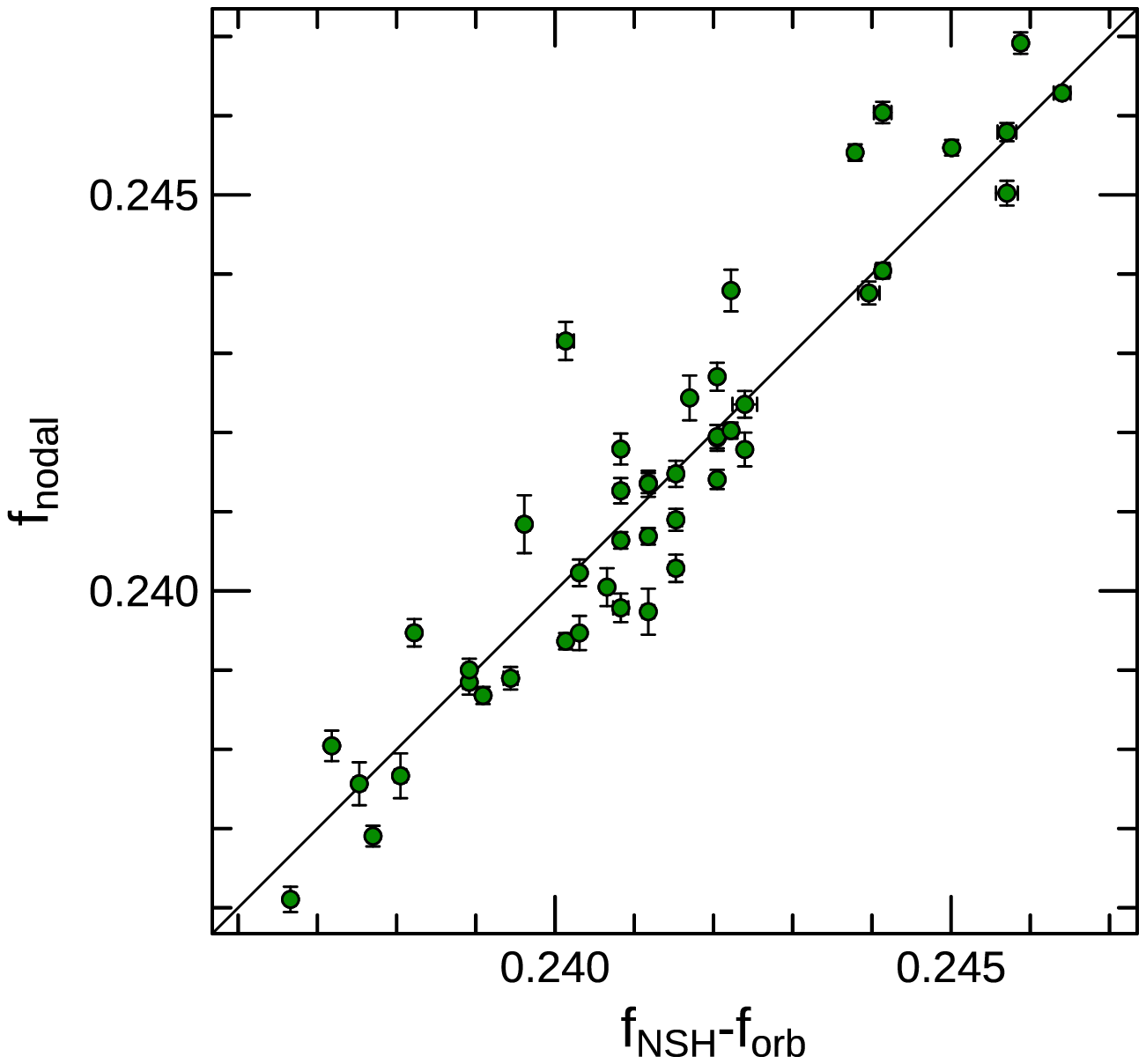}
\end{center}
\caption{$f_{\rm nodal}$ vs.~$f_{\rm NSH} - f_{\rm orb}$ displayed in the third panel of Fig.~\ref{4days}.  Ideally the points are on the solid line.  }
\label{porb-conf}
\end{figure}

We here derive the frequency variation of super-orbital 
modulations during BJD 2455296--2456312 by PDM.  
The effects of negative superhumps and orbital signals are 
removed in advance from the original light curves by using 
the results obtained before and the long-term overall trend 
of the processed light curves is subtracted by LOWESS 
(see subsection 2.2).  
We display an example of this process in Figure \ref{24cycle} 
for cycle 24 of the IW And-type phenomenon, which is 
defined in Figure \ref{ampfreq}.  
The top light curve (shown by black dots) in Figure \ref{24cycle} 
is the original light curve, and the second one (gray dots) is 
the light curve where the effects of negative superhumps 
and orbital signals are removed.  
The third one (orange dots) is the normalized light curve 
in which the long-term overall trend is subtracted.
The bottom one is the resultant one where all 
three effects: negative superhumps, orbital signals, 
and super-orbital modulations are subtracted and 
we will explain it later.  
We have applied PDM to the normalized light curve with 
a 120-d window and repeated it by shifting it by 20 d, 
as we did in subsection 3.2.2, and have derived the frequency 
of super-orbital modulations.  
We also have derived the frequency variation of negative superhumps 
by PDM with the same window size of 120-d and the same time 
step of 20-d.  Results for $f_{\rm sup}$ and 
$f_{\rm NSH} - f_{\rm orb}$ are exhibited as the triangles and 
circles in the third panel of Figure \ref{4days}, 
respectively.  Here we use the orbital frequency, 
$f_{\rm orb}$ = 3.929128 d$^{-1}$, estimated in subsection 3.3.  
We see that these two frequencies vary in unison, 
thus confirming their common origin.  

To put it in another way, we can determine the orbital 
frequency from these two observed frequencies ($f_{\rm sup}$ 
and $f_{\rm NSH}$) by assuming it as an unknown with 
the least-squares method.     
The result is displayed in Figure \ref{porb-conf} where 
the best-fit orbital period is determined to be 0.254514(7)~d. 
This value is consistent within its 1$\sigma$ error with 
the orbital period estimated in subsection 3.3.  

The period of the super-orbital modulations increases or 
decreases by $\sim$0.1~d on timescales of a few hundred days, 
which means that the radial mass distribution and/or the disk 
radius varies on long timescales according to equation 
(\ref{nu_nPR}).  
The $O-C$ curve of super-orbital modulations is displayed 
in the top panel of Figure \ref{4days} and the $O-C$ values 
are summarized in Table E7.  
The moderate variation in the $O-C$ curve reflects 
the variation of $f_{\rm sup}$.

\subsubsection{Amplitude and profile variations}

\begin{figure}[htb]
\begin{center}
\FigureFile(80mm, 50mm){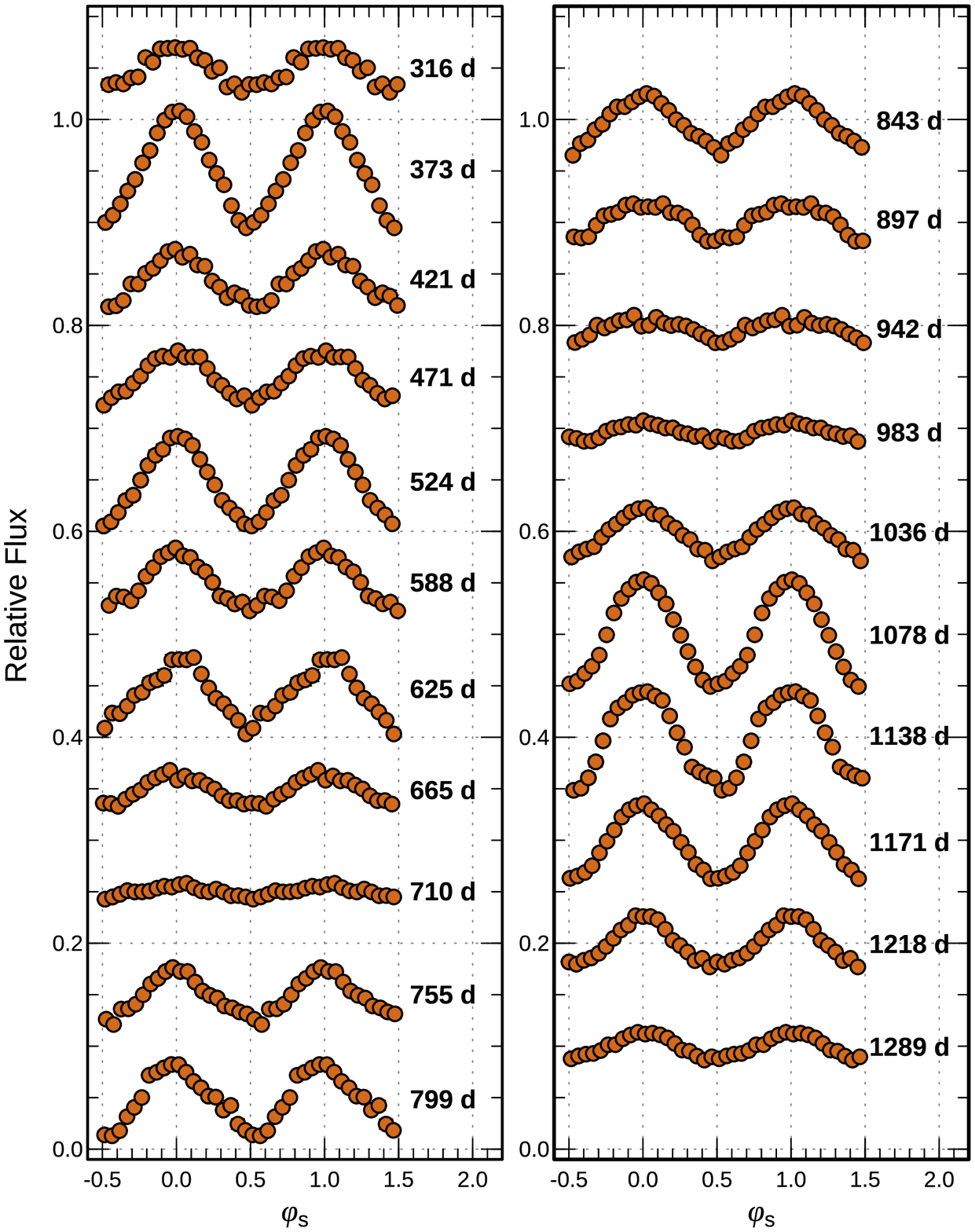}
\end{center}
\caption{Time evolution of profiles of super-orbital modulations in KIC 9406652.  We here fold the light curves normalized by the averaged flux.  The window size is $\sim$30--60 d.  The number at the right side of each profile is the date at the center of each window.  Some offsets are added to each profile for visibility in the vertical direction.  
}
\label{4days-profile}
\end{figure}

We here discuss the amplitude variation of super-orbital 
modulations.  
As shown in Figure \ref{24cycle}, we have subtracted 
the long-term overall trend of the light curves for data 
analyses; however, it is thus not easy to discern 
an individual cycle of the super-orbital modulation from 
the intrinsic variation of the disk, because the time scales 
of these two phenomena are sometimes very similar, 
particularly during quasi-standstills.  
The intrinsic variation could amplify and/or weaken 
super-orbital modulations.  
In estimating the amplitude of super-orbital modulations, 
we remove a part of data in which the intrinsic variation 
is violent (mainly around brightening) like the data around 
1040--1060~d in Figure \ref{24cycle} since we could not 
subtract it well.  

When deriving the time variation of the semi-amplitude of 
super-orbital modulations, 
we first divide the data during BJD 2455296--2456312 into 21 
intervals having $\sim$30--60-d lengths to lower the error.  
We then determine the period of super-orbital modulations, 
$P_{\rm sup} = 1/f_{\rm sup}$, for each interval from 
the frequency of negative superhumps, $f_{\rm NSH}$, 
by using the relation $f_{\rm sup} = f_{\rm NSH} - 
f_{\rm orb}$.  
This is because the direct determination of $P_{\rm sup}$ 
around 4 days is difficult for such a short interval.  
We next create a phase-averaged profile in each time interval 
by folding the normalized light curve with the period 
that we have derived.  
Finally, we fit each profile with the equation 
$a \cos (2 \pi (\varphi - b)) + 1$, where $\varphi$ is 
the phase of super-orbital modulations, since 
super-orbital light modulations resulting from 
the variation in the projected area of the tilted disk 
are supposed to have the sinusoidal waveform (see, 
equation (\ref{precess-disk-flux}) in the next subsection).
Here $a$ and $b$ are unknown constants to be fitted to 
the light curve, and $a$ stands for the semi-amplitude of 
super-orbital modulations and $b$ represents the epoch of 
its light maximum, respectively.  
The results of the regressions are summarized in Table E8.  
The phase-averaged profile of super-orbital modulations 
and their semi-amplitude in each interval are exhibited 
in Figure \ref{4days-profile} and the fourth panel of 
Figure \ref{4days}, respectively.  
We now introduce the phase of the nodal precession, 
$\varphi_{\rm s}$, in such a way that the phase zero 
corresponds to the light maximum of super-orbital modulations, 
i.e., $\varphi_{\rm s} = \varphi - b$, which is counted 
clockwise in Figure \ref{ellip-config} because the disk 
precesses retrogradely, i.e., $\nu_{\rm nPR}(t) = 
- f_{\rm sup}$.  
We have confirmed just in case that the semi-amplitude of 
periodic light variations having the super-orbital period 
in the time intervals during which negative superhumps 
were not clearly observed was less than the smallest one 
during the interval when negative superhumps were clearly 
detected.  

We see from the fourth panel of Figure \ref{4days} and 
Figure \ref{4days-profile} that the semi-amplitude varies 
quite rapidly.  
As expected, some of the profiles tend to have sinusoidal 
shapes when their amplitudes are large (i.e., at 799~d, 1078~d, 
and 1138~d in Figure \ref{4days-profile} where the dates are 
counted from BJD 2455000).  
On the other hand, some of them greatly deviate from 
the sinusoidal curve when their amplitudes are small 
(i.e., at 665~d, 710~d, 895~d, 938~d, 981~d, and 1289~d 
in the same figure). 
It would be difficult to completely remove the intrinsic 
variation of the accretion disk from super-orbital 
light variations, as expected.  
\textcolor{black}{
Although the estimate of the parameter $b$ differs 
among 21 intervals (see Table E8), this difference does not 
necessarily indicate that the orientation of the disk tilt 
is reset to new values with each cycle of the IW And-type 
phenomenon.  
This is because many factors such as the data gap, 
the remnant of the intrinsic variation of the accretion disk, 
and the difference in the period with which each set of 
light curves is folded could affect the results.  }

Now we can approximately recover the light curve  
purely originating from the accretion events 
by subtracting the three effects: 
\textcolor{black}{negative superhumps}, orbital signals, and super-orbital 
modulations, from the original light curves.  
We show the corrected light curve, in which all those 
three effects, are removed in the second panel of 
Figure \ref{4days} and also in the bottom light curve 
of Figure \ref{24cycle}.  
We see that the oscillatory variations still remain 
in quasi-standstills after the subtraction of super-orbital 
modulations: a common feature of IW And-type DNe 
(e.g, \cite{szk13iwandv513cas,kat20imeri}).

\subsubsection{Estimation of the tilt angle of the disk}

Finally, we estimate the tilt angle of the disk by using 
the semi-amplitude variation.  If we write the angle, 
which the line of sight and the normal vector of the flat 
disk make, as $\alpha$, the disk flux from the observer 
is then written by $F_{\rm disk} \cos \alpha$, where 
$F_{\rm disk}$ is the disk flux seen from pole-on.  
Let us consider that the disk is tilted out of the orbital plane 
and its normal line rotates around the normal of the binary 
orbital plane retrogradely, i.e., retrograde precession of 
the tilted disk.
The tilt angle, $\theta$, is defined by an angle between 
the normal of the orbital plane and the normal of 
the tilted disk. 
We further write the azimuthal angle of the tilted disk 
at a certain instance by $\phi$, where $\phi$ is measured 
from the direction of the line of sight. 
We note here that the azimuthal angle of the tilted disk, 
$\phi$, and the phase of super-orbital modulations, 
$\varphi_{\rm s}$, are related to each other 
by $\phi = 2 \pi \varphi_{\rm s}$.  

We consider a spherical triangle made by the three directions, 
the normal of the orbital plane, the line of sight,
and the normal of the tilted disk. 
By applying the cosine law of the spherical trigonometry to this, 
we obtain     
\begin{eqnarray}
\cos \alpha &= \cos i \cos \theta + \sin i \sin \theta \cos \phi, \\ \nonumber
&= \cos i \cos \theta (1 + \tan i \tan \theta \cos  \phi),
\label{cos-alpha}
\end{eqnarray}
where $i$ is the inclination angle of the binary system.  
If the intrinsic flux of the disk and the flux of 
the secondary star are denoted as $F_{\rm disk}(t)$ 
and $F_2$, respectively, the observed flux is written as 
\begin{eqnarray}
\label{precess-disk-flux}
& F_{\rm disk}(t) \cos \alpha+F_2 = \\ \nonumber
& (F_{\rm disk}(t) \cos i \cos \theta + F_2) 
(1 + \left(\frac{\tan i \tan \theta}{1+\frac{F_2}{F_{\rm disk}(t)\cos i \cos \theta}} \right) \cos (2 \pi \varphi_{\rm S})), 
\end{eqnarray}
where the time dependence of the intrinsic disk flux is taken 
into account by $F_{\rm disk}(t)$.  
We have normalized the light curves by $F_{\rm disk}(t) 
\cos i \cos \theta + F_2$ and the semi-amplitude $a$ 
introduced in the previous subsection corresponds to 
$\tan i \tan \theta / 
(1+\frac{F_2}{F_{\rm disk}(t)\cos i \cos \theta})$.  
According to \citet{gie13j1922}, we set $i = 50$~deg 
and used $0.06$ for $F_2 / F_{\rm disk}(t)$ as in 
subsection 3.3.\footnote{Actually, $F_2 / F_{\rm disk}(t)$ 
depends on the time variation of the disk luminosity, 
but we here neglect that variation as did in Sec.~3.3.}  
We can now calculate the tilt angle.  

The estimated tilt angle is given in the bottom panel of 
Figure \ref{4days}.  
It is less than 3~deg in KIC 9406652 in the case of 
a 50-deg inclination.  
\citet{sma09negativeSH} estimated the tilt angles of several CVs 
showing super-orbital modulations by using basically the same 
method and obtained the tilt angles around 3--4~deg on average.  
Our results seem to be consistent with his results.  
Besides, we have already used the evidence of a relatively 
low tilt angle to interpret the light curve profile of 
negative superhumps in subsection 3.2.3 (see also the left panel 
of Figure \ref{profile1}). 
Although there is a possibility that the tilt angle in KIC 9406652 
is underestimated because of the intrinsic variation of the disk 
as mentioned in subsection 3.4.2, the tilt angle unlikely exceeds 
6~deg when considering the typical thickness of the disk outer edge 
($h/r \sim 0.1$ at the outer disk edge, where $h$ stands for 
the scale height; \cite{sma92CVdisk}).  

From Figure \ref{4days}, we find three peaks in the tilt angle 
around 370~d, 520~d, and 1100~d together with a local maximum around 
800~d where dates are counted from BJD 2455000. 
When comparing this with the third panel of Figure \ref{ampfreq}, 
we see a weak correlation between the tilt angle and the amplitude 
of negative superhumps in the sense that the amplitude of 
negative superhumps is larger when the tilt angle is higher.  
We also find a weak correlation between the tilt angle and the cycle 
length of the IW And-type phenomenon when both negative superhumps 
and super-orbital modulations were clearly detected, 
in the sense that the duration of quasi-standstills tends to 
become longer as the tilt angle becomes higher, by comparing 
the second and bottom panels of Figure \ref{4days}.

\subsection{Irradiation of the secondary star and the orientation of the tilted disk}

We have shown in subsection 3.3 that the irradiation of 
the secondary plays a major role in the orbital light curve 
when the disk is tilted, in particular when the system is 
in the brightening stage (see Figure \ref{orb2}).  
In this subsection, we examine \textcolor{black}{how the orbital 
light curve looks if} we divide the observational data with respect 
to the phase of the super-orbital modulation $\varphi_{\rm s}$, 
i.e., with the orientation of the tilted disk to the line of sight.  
The data used here are restricted to those in the brightening stage 
during BJD 2455296--2456312 to focus on the irradiation effect.  
We have already determined the phase of the super-orbital modulation 
in subsection 3.4.2 and here divide the data into 8 short 
intervals with respect to $\varphi_{\rm s}$:
$\varphi_{\rm s} \sim 0.0$ (with $-$0.0625--0.0625), 0.125 
(with 0.0625--0.1875), 0.25 (with 0.1875--0.3125), 
0.375 (with 0.3125--0.4375), 0.5 (with 0.4375--0.5625), 0.625 
(with 0.5625--0.6875), 0.75 (with 0.6875--0.8125), and 0.875 
(with 0.8125--0.9375), respectively.  
We then construct 8 orbital light curves by collecting 
the data corresponding to each phase interval.  

\begin{figure}[htb]
\begin{center}
\FigureFile(80mm, 50mm){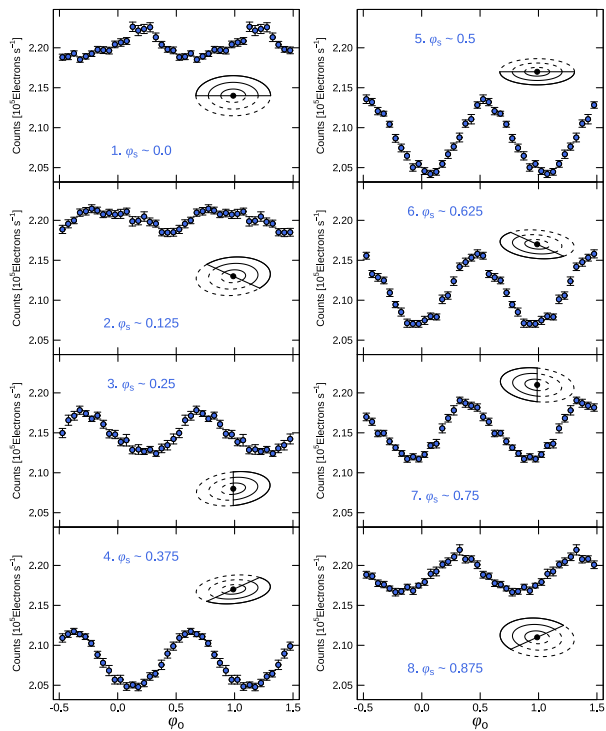}
\end{center}
\caption{Eight types of orbital profiles from the data limited for brightening during BJD 2455296--2456312.  The inset of each figure shows the orientation of the tilted disk where the central WD is marked by the black dot and the diametric line passing the WD is the nodal line of the tilted disk.  The solid and dashed lines represent the equatorial plane of the tilted disk above and below the orbital plane of the binary system, respectively.  Here, $\varphi_{\rm s}$ and $\varphi_{\rm o}$ stand for the super-orbital phase and the orbital phase, respectively, as in Figure \ref{ellip-config}.  }
\label{orb8b}
\end{figure}

Figure \ref{orb8b} exhibits the results, and we show 
the orientation of the tilted disk in each figure.  
The amplitude of the orbital light curve is the largest 
around $\varphi_{\rm s} = 0.5$ while it is the smallest 
around $\varphi_{\rm s} = 0.0$.  
This is easily understood: at $\varphi_{\rm s} \sim 0.5$, 
which corresponds to the right top panel of Figure \ref{orb8b}, 
the near side of the tilted disk is above the orbital plane 
while the far side is below the orbital plane, and hence, 
the tilted disk and the central WD then irradiate the upper face 
of the secondary most strongly at the orbital phase 
$\varphi_{\rm o} = 0.5$ and they irradiate its lower face 
most strongly at $\varphi_{\rm o} = 0.0$, 
as the secondary star goes around the tilted disk.
Here the upper/lower face of the secondary star means 
the surface of the secondary above/below the orbital plane, 
respectively.  If the orbital inclination is smaller than 50 deg, 
the observer sees mostly the irradiated upper face of the secondary 
around $\varphi_{\rm o} = 0.5$ (see also Figure \ref{ellip-config}) 
and thus the effect of irradiation of the secondary is maximal.  
The orbital profile at $\varphi_{\rm s} \sim 0.5$ shows 
maximum light at $\varphi_{\rm o} = 0.5$ and takes the largest 
amplitude.  
On the other hand, in the opposite case at $\varphi_{\rm s} \sim 0.0$ 
corresponding to the left top panel, the situation is just 
opposite to the case mentioned above, the upper face of 
the secondary is not irradiated well by the tilted disk 
around $\varphi_{\rm o} = 0.5$ and the irradiated upper face of 
the secondary is hidden around $\varphi_{\rm o} = 0.0$ from 
the observer.  
The effect of irradiation of the secondary is thus minimal 
at $\varphi_{\rm s} = 0.0$.  

We see furthermore from Figure \ref{orb8b} 
that the orbital phase of the light maximum moves to earlier 
phases as the phase of the super-orbital period, $\varphi_{\rm s}$, 
advances.  
This means that the direction of the nodal precession of 
the tilted disk is retrograde as expected from the ``negative'' 
nature of negative superhumps.  
The variation in amplitudes and the shift of peak phases 
thus give us direct evidence that the disk is actually 
tilted out of the orbital plane and precesses retrogradely 
as the secondary star acts as a reflecting mirror for 
the irradiation by the tilted disk and the WD.  
\textcolor{black}{
Also, the irradiation on the surface of the secondary star 
could affect its spectrum.  
The spectrum of the secondary star may show a higher-temperature 
component during the brightening stage, which could change 
with orbital and super-orbital phases.  }

\section{Discussion}

\subsection{Time evolution of the radius and the radial mass distribution of the tilted disk}

As shown in subsection 3.2.2, the frequency of 
negative superhumps exhibited a characteristic variation 
in one cycle of the IW And-type phenomenon of KIC 9406652: 
a rapid decrease at the beginning of brightening followed 
by a small increase and a gradual increase during 
quasi-standstills (see the bottom panels of 
Figures \ref{ampfreq} and \ref{pick}).  
We interpret this frequency variation, introducing some results 
in \citet{kim20tiltdiskmodel}.  
They predicted the variation in the nodal precession rate, 
$\nu_{\rm nPR}$, given in equation (\ref{nu_nPR}), which is 
directly related to the frequency of negative superhumps 
by equation (\ref{nu_nSH}), via their numerical simulations 
of the thermal-viscous instability in the non-tilted and 
tilted disks.\footnote{Here the non-tilted 
disk model means the model in the limiting case of 
the lowest tilt angle.  }

We first discuss the frequency variation at the beginning 
of the brightening.  
Generally, the disk expands at the beginning of a (small) 
outburst \citep{ich92diskradius} and then the frequency of 
negative superhump rapidly tries to increase.  
We see from the bottom panels of Fig.~10 and Fig.~12 in 
\citet{kim20tiltdiskmodel} that there are two patterns 
in its time evolution after the initial variation: 
a continuous increase and a rapid decrease.  
The latter occurs if a huge amount of mass accumulates 
at the outer disk and the disk reaches the tidal truncation 
radius.  
\textcolor{black}{
In this case, the angular-momentum loss by the tidal torque 
exerted by the secondary star is considered to keep being 
efficient for a while and a huge amount of mass accumulated 
in the outer disk accretes on the WD.  
In other words, the disk becomes more or less near 
the standard disk with $\Sigma \propto r^{-3/4}$ 
during outburst from the quiescent disk with $\Sigma$ 
having a positive slope with $r$.  
According to the second row of equation (\ref{nu_nSH}), 
the frequency of negative superhumps depends only on 
the weight of the disk mass in the radial direction, i.e., 
$\Sigma (r)$, if the disk radius is fixed to the tidal 
truncation radius.  
If the weight of the mass in the inner disk increases, 
the frequency decreases.  }
The observed rapid drop, therefore, seems to represent 
the rapid change of the surface-density distribution, 
as actually predicted in the tilted-disk model in 
\citet{kim20tiltdiskmodel}.  
This interpretation would be consistent with the spectral 
energy distribution observed at the brightening stage 
in IW And stars, 
which was close to that of the standard disk 
\citep{szk13iwandv513cas}.  

On the other hand, the observed gradual increase in 
the frequency during the quasi-standstill is against 
the prediction by \citet{kim20tiltdiskmodel}.  
In their tilted disk models, the outer part of the disk 
is supposed to be in the cool state during the quasi-standstill.  
The addition of matter at the disk edge by the gas stream 
(having less specific angular momentum than that of the disk edge) 
causes contraction of the disk radius, which leads to 
a gradual decrease in the nodal precession rate, i.e., 
the frequency of negative superhumps.  
The gradual increase in the negative-superhump frequency 
in the observations thus suggests the gradual increase 
in the disk radius.  
The disk expansion during standstills was already suggested 
in NY Ser, an SU UMa-type star by \citet{kat19nyser}.  
This object entered a standstill in 2018 after frequent normal 
outbursts, which were terminated by a superoutburst.  
In order for a superoutburst to start, the disk radius must reach 
the 3:1 resonance radius at the beginning of the superoutburst 
\citep{osa89suuma}, and thus the disk radius should have 
gradually increased during standstills.  
\citet{kat19iwand} suggested that the same situation 
could occur in the case of IW And-type stars as well.  

It seems difficult to realize a gradual increase 
in the disk radius during the quasi-standstill.  
The outer part of the disk must be in the hot state 
in order for the disk radius to expand.  
However, if the disk is in the hot state under a condition of 
weak tidal torques (i.e., its radius is below the tidal 
truncation radius), the disk would then expand rapidly and 
it could not stay in the quasi-steady state for a long while.  
The gradual increase in the negative-superhump frequency, 
which is suggested by observations during the quasi-standstill, 
poses a new question about the nature of the quasi-standstill 
in IW And-type stars, to which we do not have any immediate 
answers.

\subsection{Examination of the models for the IW And-type stars}

As mentioned in the introduction, two models have been proposed 
so far to explain the characteristic light variation 
in IW And stars.  
We first examine the mass-transfer-burst model proposed 
by \citet{ham14zcam} by using our results.  
Although they did not discuss the disk radius variation 
in their paper, the mass transfer burst from the secondary 
would lead to a sudden contraction of the disk radius 
(see \cite{ich92diskradius}).  This phenomenon may seem 
to be consistent with the observed sudden drop 
in the frequency of negative superhumps, which occurred 
at the beginning of brightening.  
However, their model is inconsistent with our present results 
in other aspects.  
As demonstrated in subsection 3.2.3, the light source of 
negative superhumps is the bright spot of the gas stream 
sweeping on the surface of the tilted disk.  
If 10-times increases in mass transfer rate are needed to 
reproduce the brightening as predicted by \citet{ham14zcam}, 
the amplitude of negative superhumps must drastically 
increase at the brightening as shown in Fig.~1 of 
\citet{woo07negSH}.  
This expectation is inconsistent with the observational fact 
that the amplitude of negative superhumps did not change 
at the brightening phase of the IW And-type light cycle but 
rather it became a little smaller (see subsection 3.2.3).  
Furthermore, no enhancement of the orbital hump was observed 
either at the brightening.  
These results seem to rule out the enhanced mass transfer model 
by \citet{ham14zcam}: the same argument for the disk instability 
model in the ordinary DN outbursts over the mass-transfer-burst model.

Next, we examine the other model proposed by 
\citet{kim20tiltdiskmodel}, in which the thermal-viscous 
instability in the tilted disk was studied in 
the framework of the disk instability model.  
Some of the observational results seem to be supportive of 
this model.  
The observed amplitude variation of negative superhumps 
(see the third panel of Figure \ref{ampfreq}) 
during one cycle of the IW And-type phenomenon seems to be 
consistent with the assumption that the mass transfer rate 
stays more or less constant in their model.  
Moreover, the weak correlation between the tilt angle and 
the cycle length of the IW And-type phenomenon seems to be 
consistent with their model, as the cycle interval is 
longer when the tilt angle is higher (see also subsection 3.4.3).  
The rapid decrease in the negative-superhump frequency 
at the brightening phase was also predicted by them 
(see subsection 4.1).  

On the other hand, some other results are not supportive of 
their model.  
As discussed in subsection 4.1, the gradual increase in 
the observational frequency variation of negative superhumps 
during quasi-standstills cannot be reproduced by their model.  
Also, it has turned out that the simplified assumption in which 
the gas stream takes a single-particle trajectory in their model 
is not realistic to explain the \textcolor{black}{light curve} of negative superhumps, 
as described in subsection 3.2.3.  
They calculated various mass supply patterns 
from the secondary under the simplified assumption and 
reproduced by their numerical simulations 
a cyclic light curve similar to the IW And-type phenomenon only 
in the highly-tilted disk model, 
while the observations suggest a very low tilt angle and 
the IW And-type phenomenon was observed even when 
negative superhumps were very weak or disappeared in 
KIC 9406652 (see subsections 3.2.1 and 3.4.3).  
Actually, IM Eri, another IW And star, did not show clearly 
negative superhumps during its IW And-type phenomenon 
\citep{kat20imeri}.  
We have to modify the calculation of the mass supply pattern 
by taking into account the gas-stream overflow and confirm 
whether the IW And-type phenomenon and the gradual increase 
of the negative-superhump frequency during quasi-standstills 
can be reproduced in the non-tilted or slightly-tilted disk.  
There is a possibility that the most essential condition for 
the IW And-type phenomenon may be the overflow of 
the gas stream at the disk rim, reaching the inner part of 
the disk both in the case of the non-tilted disk and of 
the tilted disk alike \citep{hes99streamoutflow,sch98streamoverflow}.

\section{Summary}

We re-investigated the {\it Kepler} data of KIC 9406652 
and performed detailed timing analyses of negative superhumps, 
orbital signals, and super-orbital modulations.  
Our main results are summarized as follows.  
\begin{enumerate}
\item
The emergence of negative superhumps was not necessarily 
correlated with the IW And-type phenomenon (see subsections 
3.1 and 3.2.1), which implies that the IW And-type phenomenon 
may not be directly related to the disk tilt.  
\item 
The frequency of negative superhumps showed cyclic variations 
corresponding to each cycle of IW And-type phenomena: 
a rapid decrease during the brightening and a gradual increase 
during the quasi-standstill (see subsection 3.2.2).  
This variation would represent the rapid change of 
the surface density distribution during the brightening 
and the gradual expansion of the disk during 
the quasi-standstill (see subsection 4.1).  
Thus the investigation of frequency variations of 
negative superhumps will provide an important new tool 
for understanding the variation of the disk radius and/or 
the radial mass distribution in a tilted accretion disk.  
\item 
The amplitudes of the negative superhumps did not vary 
with time so much in the flux scale, even though 
the intrinsic disk luminosity varies greatly during 
one cycle of the IW And-type phenomenon (see subsection 3.2.3).  
This suggests that negative superhumps are caused by 
the variable dissipation of the bright spot as the gas stream 
sweeps the tilted disk surface with the negative superhump cycle.  
\item
The \textcolor{black}{light curve} of the negative superhumps was single-peaked 
and it showed a great variety (see subsection 3.2.3).  
This implies that a part of the gas stream always overflows 
the disk edge and other parts collide with the disk edge 
in the slightly tilted disk.  
\item 
\textcolor{black}{The orbital signal consists of} the three components: 
the irradiation of the secondary by the tilted disk and 
the WD, the orbital hump formed at the disk rim by 
the gas stream, and the ellipsoidal modulation 
(see subsection 3.3).  
The irradiation effect was dominant when the disk 
became bright, while the orbital hump dominated 
when the system was in the dip and/or when the tilt angle of 
the disk was almost non-tilted.  
The contribution of the ellipsoidal modulation is 
very small, only a few percent to the total orbital signal.  
\item 
The frequency variation of super-orbital modulation was 
consistent with that of negative superhumps, which means 
that both of the phenomena have the same origin: 
the tilted disk (see subsection 3.4.1).
\item
The tilt angle was estimated to be less than 3 deg 
from the time variation of the semi-amplitude 
of super-orbital modulations, which are consistent with 
the amplitude and profile variations of negative superhumps 
(see subsection 3.4.3).  
\item
The orbital \textcolor{black}{light curve}, particularly in the brightening stage, 
varied greatly with the phase of super-orbital modulation, 
i.e., with the orientation of the tilted disk to the observer 
(see subsection 3.5).  
Its amplitude was the largest at the minimum of super-orbital 
modulations and the smallest at the opposite phase.  
The light maximum of the orbital light curve shifted to 
an earlier phase as the phase of the super-orbital modulation 
advances.  
These phenomena are understood as the upper face of 
the secondary star is irradiated by the tilted disk: 
clear evidence that the disk was tilted out of 
the orbital plane of the binary system and it precessed 
`retrogradely'.  
\item 
We have evaluated the recently proposed two models for 
IW And stars by the results of our analyses (see subsection 4.2).  
Any supportive results have not been obtained for the model 
proposed by \citet{ham14zcam}.  
On the other hand, a part of the frequency variation of 
negative superhumps does not support the model proposed by 
\citet{kim20tiltdiskmodel}, while some of our results: 
the amplitude variation of negative superhumps and 
the correlation between the tilt angle and the cycle 
length of the IW And-type phenomenon seem to support 
this model.  
There might be a possibility that the extension of 
this model can reproduce the IW And-type phenomenon 
better if we take into account the gas-stream overflow.  
\end {enumerate}

\section*{Acknowledgements}

This work was financially supported by the Grant-in-Aid for 
JSPS Fellows for young researchers (M.~Kimura).  
M.~Kimura acknowledges support by the Special Postdoctoral Researchers
Program at RIKEN.  
The numerical code that we used in this paper is provided
by Izumi Hachisu.
\textcolor{black}{We thank the anonymous referee for helpful comments.  }

\section*{Supporting information}

Additional supporting information can be found in the online version 
of this article:
supplementary tables E1--E8 and supplementary figures E1--E28.

\newcommand{\noop}[1]{}


\end{document}